\documentclass[12pt]{article}

\usepackage{scicite}

\usepackage{times}

\usepackage{amsmath}
\usepackage{amsfonts}
\usepackage{amssymb}
\usepackage{graphicx}

\usepackage{graphicx}
\usepackage{psfrag}
\usepackage{subfigure}
\usepackage{epsfig}
\usepackage{calc}
\usepackage{bm}
\usepackage{amsmath}
\usepackage{ulem}
\usepackage{color}
\usepackage{nicefrac}
\usepackage{caption}
\graphicspath{{figures/}}
\usepackage{hyperref}
\usepackage{xr}
\newenvironment{sciabstract}{%
\begin{quote} \bf}
{\end{quote}}

\title{Filamentous Active Matter:\\
 Band Formation, Bending, Buckling, and Defects}

\author{Gerard A. Vliegenthart$^{\ast}$, Arvind Ravichandran, Marisol Ripoll,\\
 Thorsten  Auth, Gerhard Gompper\\
\\
\normalsize{Theoretical Soft Matter and Biophysics},\\
\normalsize{Institute of Complex Systems and Institute for Advanced Simulation,}\\
\normalsize{Forschungszentrum J\"ulich, 52425 J\"ulich, Germany}\\
\normalsize{$^\ast$To whom correspondence should be addressed; E-mail:  g.vliegenthart@fz-juelich.de.}
}

\date{}

\begin{document} 


\baselineskip24pt


\maketitle



\begin{sciabstract}
Motor proteins drive persistent motion and self-organisation of cytoskeletal filaments.
However, state-of-the-art microscopy techniques and continuum modelling approaches focus on large length and time scales.
Here, we perform component-based computer simulations of polar filaments and molecular motors linking microscopic interactions and activity to self-organisation and dynamics from the two-filament level up to the mesoscopic domain level.
Dynamic filament crosslinking and sliding, and excluded-volume interactions promote formation of bundles at small densities, and of active polar nematics at high densities. A buckling-type instability sets the size of polar domains and the density of topological defects.
We predict a universal scaling of the active diffusion coefficient and the domain size with activity, and its dependence on parameters like motor concentration and filament persistence length.
Our results provide a microscopic understanding of cytoplasmic streaming in cells and help to develop design strategies for novel engineered active materials.
\end{sciabstract}

\section*{Introduction}

The key structural and active element operating biological cells is their cytoskeleton, which to a large extent 
is composed by polar filaments dynamically interconnected by passive and active crosslinkers~\cite{KOHLER011,SCHMIDT011}. 
The cytoskeleton provides mechanical stability to cells, acts as dynamic force-generating element, and serves as a track network for active intracellular transport~\cite{FLETCHER010,LIN016}. Moreover, internal motion of the cytoplasm, 
which can be generated  by the dynamics of the cytoskeleton, secures nutrient availability and the distribution of organelles~\cite{LU016}. 
Fundamental knowledge about the relationship between cytoskeleton  structure and dynamics is 
therefore necessary to obtain a deeper understanding of cellular function and dysfunction in vivo, 
and for design and synthesis of active-gel materials, e.g., artificial 
cells~\cite{KOENDERINK011,NEEDLEMAN017}. 
In dilute suspensions of freely diffusing filaments and motors, 
the filaments self-organise into large-scale aster-like structures and vortices~\cite{URRUTIA91,NEDELEC97}. 
In contrast, in dense suspensions of filaments and motors at an oil-water interface, pioneering 
experiments~\cite{SANCHEZ012,KEBER014,HENKIN014,GUILLAMAT017} 
have revealed exciting non-equilibrium  behaviours, such as persistent spontaneous flows and 
turbulence. In particular, the formation of topological defects has been shown to be a clear 
signature of active nematics \cite{SANCHEZ012}. 

Component-based computer simulations have greatly advanced the field of passive colloidal 
systems  in the past, as well as the emerging field of active materials in recent times.  In the 
latter, non-equilibrium, energy-consuming processes drive the system, which lead to very rich 
collective behavior \cite{RAMASWAMY010,MARCHETTI013,SHELLEY015,ELGETI015}.
For example, ensembles of self-propelled rod-like particles form novel liquid-crystalline steady 
states~\cite{WENSINK012,ABKENAR013,GIOMI014,SPUDICH83,OOSAWA84}. 
Single self-propelled, semiflexible filaments form spirals, while multiple filaments cluster and 
their dynamics changes from jamming to active turbulence with increasing Peclet number \cite{DUMAN018}.  
At high densities, where the equilibrium phase of the corresponding passive system is nematic, 
continuum models predict spontaneous flows and 
turbulence~\cite{GIOMI014,GIOMI015,THAMPI013,THAMPI014,YEOMANS016,HEMINGWAY016}.
Computational modelling has also been used to study cellular processes connected to cytoskeletal 
dynamics, like mitosis and contractility~\cite{KARSENTI06,NAZOCKDAST017}. 

Important and challenging questions for filament-motor systems include the identification of the 
roles of hydrodynamic interactions, filament flexibility, and motor and crosslinker properties.
While existing continuum models indicate that hydrodynamic interactions are an essential element for the formation of 
defect structures~\cite{MARCHETTI013,YEOMANS016}, this is in contrast with results obtained with 
models of apolar active elipsoids~\cite{SHI013} and self-propelled filaments~\cite{DUMAN018}, and 
experiments of vibrated granular rods~\cite{NARAYAN07}.
In these systems, defect formation and complex dynamics have been observed in the absence of hydrodynamics.
Flexibility of the filaments must also be relevant, because in simulations of mixtures of stiff rods and
motors topological defects have not been observed~\cite{BETTERTON016,RAVICHANDRAN017,RAVICHANDRAN019}.
Finally, passive cross-linkers have been demonstrated to be key to change the behavior 
of a filament-motor mixture from extensile to contractile \cite{BELMONTE017}.

In this paper, we study the emergent structures and persistent dynamics in mixtures of semiflexible 
filaments and molecular motors using Langevin Dynamics simulations. Our microscopic, filament-based 
modelling approach for dilute and concentrated systems bridges length scales from nanometer-sized molecular motors 
to micrometer-long, semiflexible filaments, and time scales from tens of microseconds for single-motor steps to 
seconds for cytoplasmic streaming \cite{HEAD011,RAVICHANDRAN017}.  
Our two-dimensional simulations correspond to experimental studies for filament-motor suspensions at oil-water 
interfaces \cite{SANCHEZ012}.  
Our calculations show that the average motor-induced force on antiparallel filaments is a robust measure 
for the activity in the system. The generated active stresses induce a buckling-type instability in initially nematic 
filament systems. At steady state, complex flow patterns 
emerge that lead to continuous creation and annihilation of topological defects, and to active Brownian particle-like 
filament diffusion at long time scales.

Our simulations are based on a model of 
semiflexible filaments of contour length $L$ and persistence length $\ell_\mathrm{p}$ in two dimensions.
Filaments consist of $n_\mathrm{s}$ beads of diameter $\sigma$ connected by stiff harmonic springs with 
rest length $a_0=L/(n_\mathrm{s}-1)$ where $a_0=\sigma$.
The  filament area fraction $\phi=n_\mathrm{f}n_\mathrm{s}\pi\sigma^2/4L_\mathrm{box}^2$ is varied by changing the number of filaments, $n_\mathrm{f}$, 
or the box size $L_\mathrm{box}$.  
Molecular motors are modelled by harmonic springs with rest length $r_0$ and spring constant $k_\mathrm{m}$, and attach 
to neighbouring filaments with rate $\Gamma_\mathrm{att}$. Motors walk in the direction of the filament 
polarity with step length $a_0$. The step rate is proportional to the probability $p_\mathrm{m}^0$ to move a motor arm, which sets the bare motor velocity 
$v_0=a_0 p_\mathrm{m}^0/\Delta t$, where $\Delta t$ is the motor time step. Motors detach when they reach the end of a filament, when they encounter a motor already bound, or when their 
length exceeds a threshold $r_\mathrm{off}$. 
In the following we use the reduced quantities: filament aspect ratio $\tilde{L}=L/\sigma$, motor-to-filament ratio 
$\tilde{n}_\mathrm{m}=n_\mathrm{m}/n_\mathrm{f}$, persistence length $\tilde{\ell}_\mathrm{p}={\ell}_\mathrm{p}/L$, motor spring constant 
$\tilde{k}_\mathrm{m}=k_\mathrm{m}r_0^2/k_\mathrm{B}T$ ($k_\mathrm{B}T$ is the thermal energy), filament friction 
$\tilde{\gamma}=\gamma/k_\mathrm{m}\delta t$ ($\delta t$ is the simulation time step), box size $\tilde{L}_\mathrm{box}=L_\mathrm{box}/L$, and time $\tilde{t}=t D_0/L^2$, 
with $D_0$ the single passive-filament translational diffusion coefficient. For details see Methods.

\section*{Results}
\subsection*{Activity drives the formation of polar domains}
\begin{figure*}
\center
\includegraphics[height=0.245\textwidth]{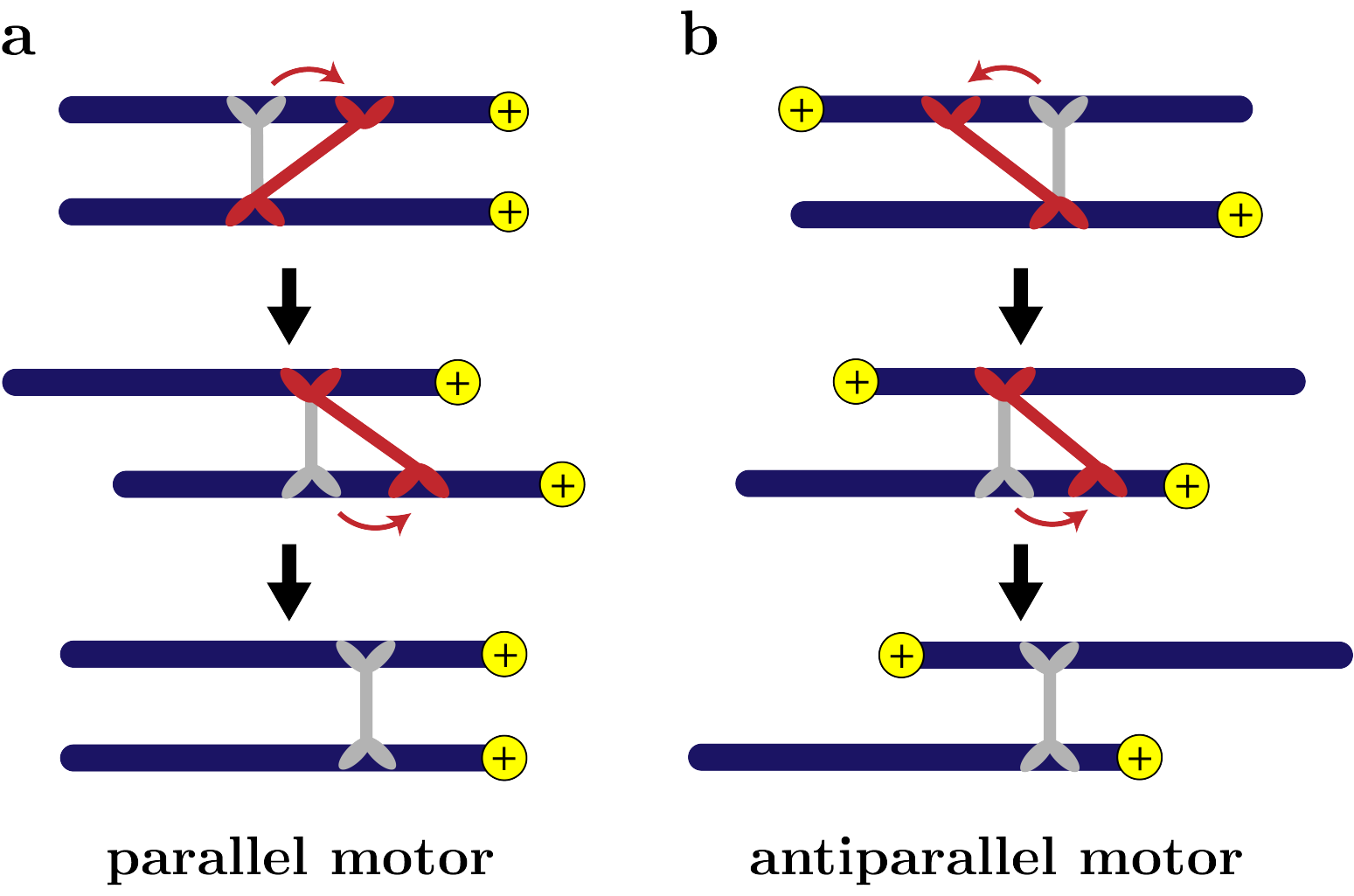}\hspace{0.2cm}
\includegraphics[height=0.05\textwidth]{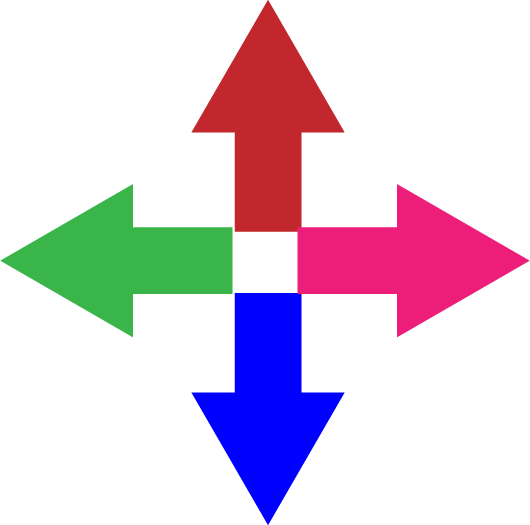}
\includegraphics[height=0.25\textwidth]{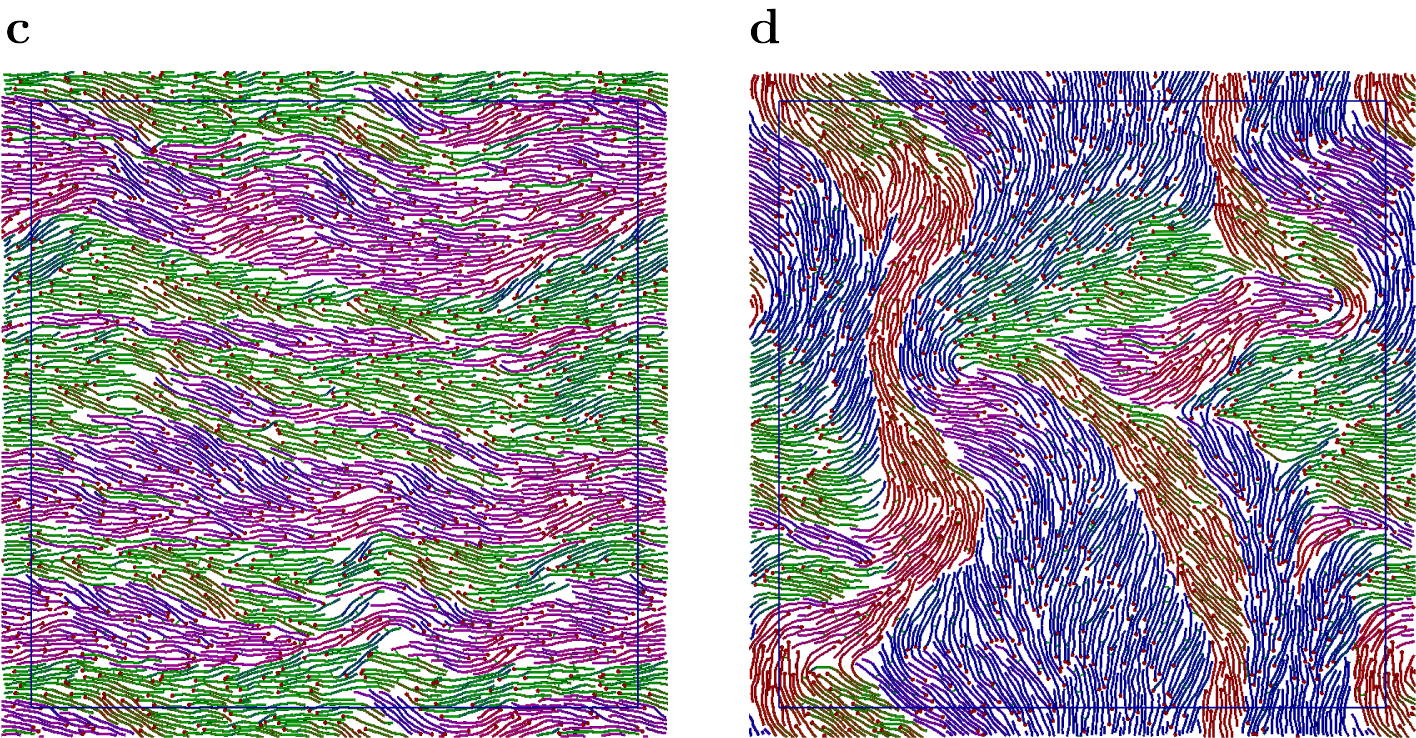}\\
\vspace*{1cm}\raisebox{-0.5\height}{\includegraphics[width=0.35\textwidth]{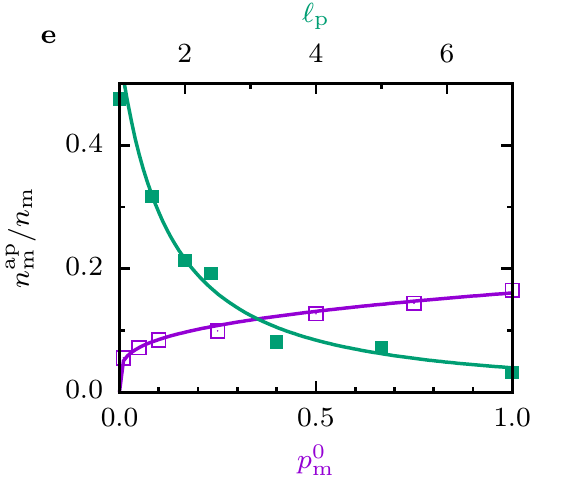}\hspace*{-0.5cm}
\includegraphics[width=0.35\textwidth]{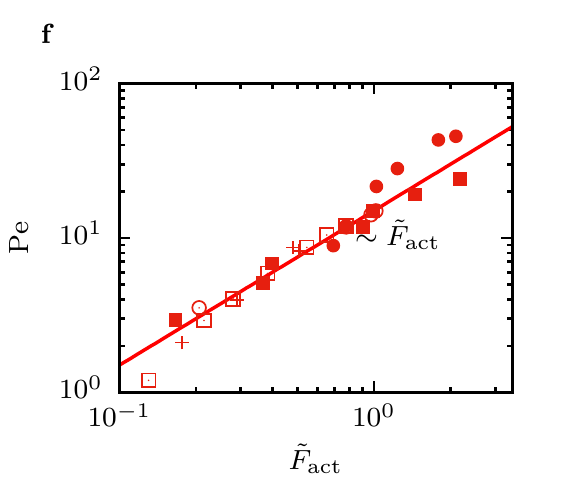}}
\hspace*{-1.cm}\scriptsize{\begin{tabular}{cllllll}
\hline \hline
& ${\tilde{\ell}_\mathrm{p}}$  & $\tilde{k}_\mathrm{m}$ & $\tilde{\gamma}$ & $p_\mathrm{m}^0$ & $\tilde{n}_\mathrm{m}$\\
\hline
$\blacksquare$	& 1-7& 12.5 & 4 & 1    & 0.89 \\
$\odot$		& 2.4   & 6.25-50 & 1-8 & 1    & 0.89 \\
{\large $\bullet$}		& 2.4 & 12.5 & 0.4-400 & 1    & 0.89 \\
$\Box$		& 2.4 & 12.5 & 4 & 0.01-1 & 0.89\\
$+$			& 3.4 & 12.5 & 4 & 1    & 0.5-3.6 \\
\hline \hline
\end{tabular}}\\
\caption{{\bf Self-organised structures and dynamics of motor-filament mixtures driven by sliding of antiparallel filaments.\\\hspace{\textwidth}}
\textbf{a,b}~Sketches of two consecutive motor steps when filaments are oriented \textbf{a}~parallel  \textbf{b}~antiparallel. 
The initially relaxed grey motor steps towards the filament polar end, it relaxes and it makes a second step on the other filament and 
relaxes again. This results in net filament motion only for antiparallel motors. \textbf{c,d}~Simulation snapshots for an initially disordered 
nematic system. The thin black box indicates the central simulation cell. \textbf{c} short-time band formation and \textbf{d} long-time disordered structures.  Filament colours indicate their orientation,  
illustrated by the colour axis.  \textbf{e}~Number of antiparallel motors as a function of the persistence length $\ell_\mathrm{p}$, and the bare 
motor velocity, measured by $p_\mathrm{m}^0$.  \textbf{f}~Peclet number as a function of the active force $\tilde{F}_\mathrm{act}$ in \protect{Eq.~(\ref{eq:Fact}}). 
The table shows the parameter combinations used in~(\textbf{e},\textbf{f}). In all cases ($\phi,\tilde{L}$)=(0.66, 20).}
\label{fig:snapshots}
\end{figure*}

The microscopic origin of filament motion is the active and dynamic crosslinking of filaments by molecular motors. Because filaments are intrinsically polar, the resulting parallel forces depend on the relative orientation of connected filaments.
If these two filaments are  polar-aligned, and two consecutive motor steps occur on the two different filaments, 
the motors induce no net filament motion, see Fig.~\ref{fig:snapshots}\textbf{a}. However, if the filaments are  anti-aligned, 
the motors get stretched and net filament motion results, see Fig.~\ref{fig:snapshots}\textbf{b}. The $n_\mathrm{m}$ motors are classified 
as $n_\mathrm{m}^\mathrm{p}$ 'parallel motors'  that connect parallel filaments in the interior of domains, and ${n_\mathrm{m}^\mathrm{ap}}$ 'anti-parallel motors' 
that dynamically crosslink and slide filaments  at the interfaces between oppositely oriented domains relative to each other. 

An initially disordered nematic suspension of filaments transforms into an "{\it active polar nematic}" when molecular motors are added.
The motor-induced sliding forces first lead to a rapid sorting of 
filaments into narrow polar bands with antiparallel alignment at the domain boundaries, see 
Fig.~\ref{fig:snapshots}\textbf{c}. In time, these polar bands coarsen. When the activity is large enough, a 
buckling instability leads to disordered configurations of polar domains with topological defects, see 
Fig.~\ref{fig:snapshots}{\textbf{d}}.  If the activity is too small or the 
filaments are too stiff, the steady state consists of stable parallel bands, see fig.~SI~1.   
Video~1  illustrates the polarity-sorting process in more detail. 
Video~2 shows the entire dynamical evolution from the initial nematic state to the stationary state.

Steady-state filament dynamics emerges from a complex interplay between various filament and motor properties. 
For example, the number of antiparallel motors is an important factor driving the filament dynamics, as well as the suspension structure. 
Figure~\ref{fig:snapshots}\textbf{e} shows that the fraction of antiparallel motors increases with decreasing filament 
persistence length $\ell_\mathrm{p}$ and with increasing bare motor velocity. As we will show later, a shorter persistence length leads to 
smaller domains and overall larger interface length, such that more antiparallel motors can be accommodated. Similarly, larger motor 
velocities lead to a higher activity and to smaller domains.  We quantify the activity by the total motor force in the 
system as
\begin{equation}
\tilde{F}_\mathrm{act}={n_\mathrm{m}^\mathrm{ap}} \tilde{k}_\mathrm{m} \left( \frac{{r_\mathrm{m}}}{r_\mathrm{0}}-1 \right) \label{eq:Fact}
\end{equation}
with the average relative extension ${r_\mathrm{m}}/r_\mathrm{0}-1$ 
of the antiparallel motors. 
In steady state, the motor force is balanced by an effective friction force with friction coefficient $\gamma_\mathrm{eff} \propto D_0^{-1}$; 
 the parallel filament velocity $v_{\parallel}$ is defined as 
${v_{\parallel}=\lim_{\tau  \to 0}\langle -{\bf p}(t) \cdot \left ({\bf r}(t+\tau)-{\bf r}(t)\right)/\tau\rangle_t}$, 
where ${\bf r}(t)$ and ${\bf p}(t)$ are the position and the unit tangent vector of the filament center, respectively, and $\tau$ is the 
lag time, see fig.~SI~2. The parallel velocity and thus the Peclet number 
$\mathrm{Pe} =v_{\parallel} L/D_0$  increases linearly with the activity (total active force) in the system, see Fig.~\ref{fig:snapshots}~\textbf{f}.
Note that the active force $\tilde{F}_\mathrm{act}$ indirectly depends on $\ell_\mathrm{p}$ and $\tilde{n}_\mathrm{m}$, thus the same magnitude can be achieved by different 
combinations of parameters. 

\subsection*{Single-filament motion and active diffusion}
\begin{figure*}
\centering
\includegraphics[width=0.35\textwidth]{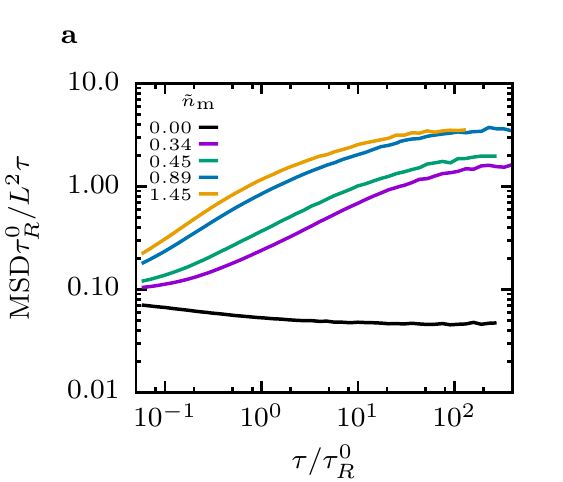}
\hspace*{-0.6cm}\includegraphics[width=0.35\textwidth]{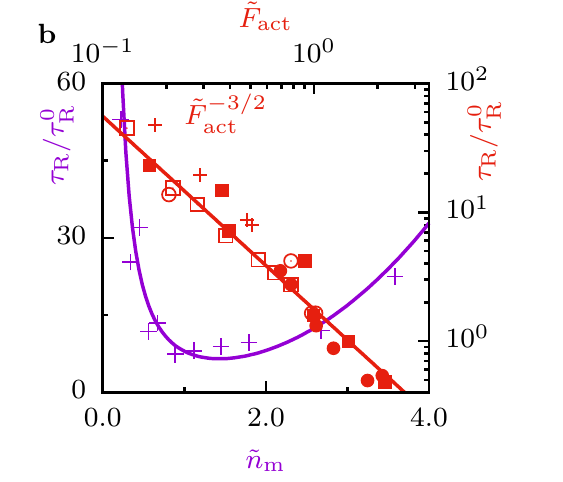}
\hspace*{-0.5cm}\includegraphics[width=0.35\textwidth]{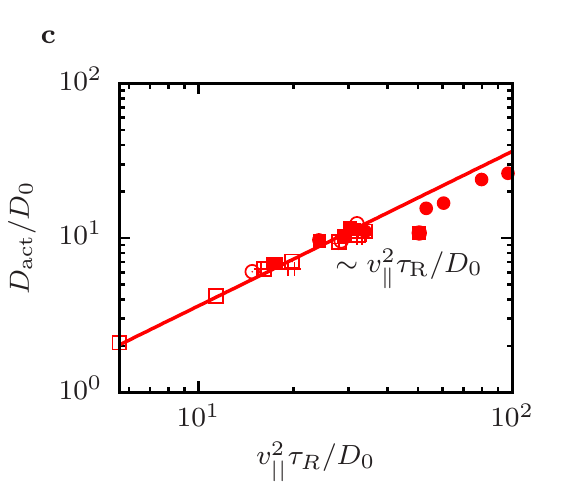}
\caption{{\bf Single-filament motion and active diffusion for various propulsion forces.\\\hspace{\textwidth}} \textbf{a} Filament mean squared displacement divided by the lag time for different motor concentrations $\tilde{n}_\mathrm{m}$. The lag time $\tau$ is 
normalised with the passive single-filament rotation time $\tau_\mathrm{R}^0$, and the MSD with the filament length $L^2$. 
The parameters are ($\phi, \tilde{\ell}_\mathrm{p}, \tilde{L}$)=(0.66, 3.4, 20).
\textbf{b} Normalised active rotation times $\tau_\mathrm{R}$ as function of $n_\mathrm{m}$ and  $\tilde{F}_\mathrm{act}$. Solid lines are a guide to the eye, 
parameters as in \protect{Fig.~\ref{fig:snapshots}\textbf{f}} with ($\phi,\tilde{L}$)=(0.66, 20). 
\textbf{c} Normalised active diffusion constants obtained from MSD-curves as a function of the filament measured $v_{\parallel}^2\tau_\mathrm{R}$, 
parameters as in \protect{Fig.~\ref{fig:snapshots}}\textbf{f} with ($\phi,\tilde{L}$)=(0.66, 20).}
\label{fig:D_act}
\end{figure*}

To characterise single filament motion in the active steady states, we
calculate the filament orientational autocorrelation functions and
filament center-of-mass mean squared displacements (MSD). 
For non-zero $\tilde{n}_\mathrm{m}$ and intermediate times, the filaments
follow essentially straight trajectories. Their ballistic motion,
with $\textrm{MSD} \propto v_{\parallel}^2 \tau^2$, is a signature of this active
persistent motion.  At long times, the filaments lose their
initial orientation due to active rotational motion, and the MSD
becomes again linear with time, characterised by a plateau in
Fig.~\ref{fig:D_act}\textbf{a}, with an active diffusion coefficient
$D_{\rm act}$ that increases with increasing $\tilde{n}_\mathrm{m}$.  The filament
orientational autocorrelation functions decay exponentially with an active
decay time $\tau_\mathrm{R}$, which also depends on $n_\mathrm{m}$.  The active rotation
time $\tau_\mathrm{R}$ is shown in Fig.~\ref{fig:D_act}\textbf{b} both as a
function of $n_\mathrm{m}$ and $F_\mathrm{act}$.  For small $n_\mathrm{m}$, increasing the number
of motors first leads to a faster motion but at $\tilde{n}_\mathrm{m} \approx 1.5$ the
motion becomes slower again because the number of antiparallel motors per filament ${n_\mathrm{m}^\mathrm{ap}}$
    does not increase anymore (see fig.~SI~3); the excess motors
increase the crosslinking density inside the polar domains which reduces the filament velocity.
The active rotation time $\tau_\mathrm{R}$ decreases with increasing $F_\mathrm{act}$
(${n_\mathrm{m}^\mathrm{ap}}$) with a power law, $\tau_\mathrm{R} \propto
F_\mathrm{act}^{\nicefrac{-3}{2}}$.  The active diffusion
coefficient, parallel velocity, and rotational correlation times for many
different parameter variations are related by $D_{\rm act} \propto
v_{\parallel}^2 \tau_\mathrm{R}$, see Fig.~\ref{fig:D_act}\textbf{c}, consistent with the theory of active Brownian
motion~\cite{HOWSE07}.

\subsection*{Buckling polar bands}
\begin{figure*}
\centering
\includegraphics[width=0.68\textwidth]{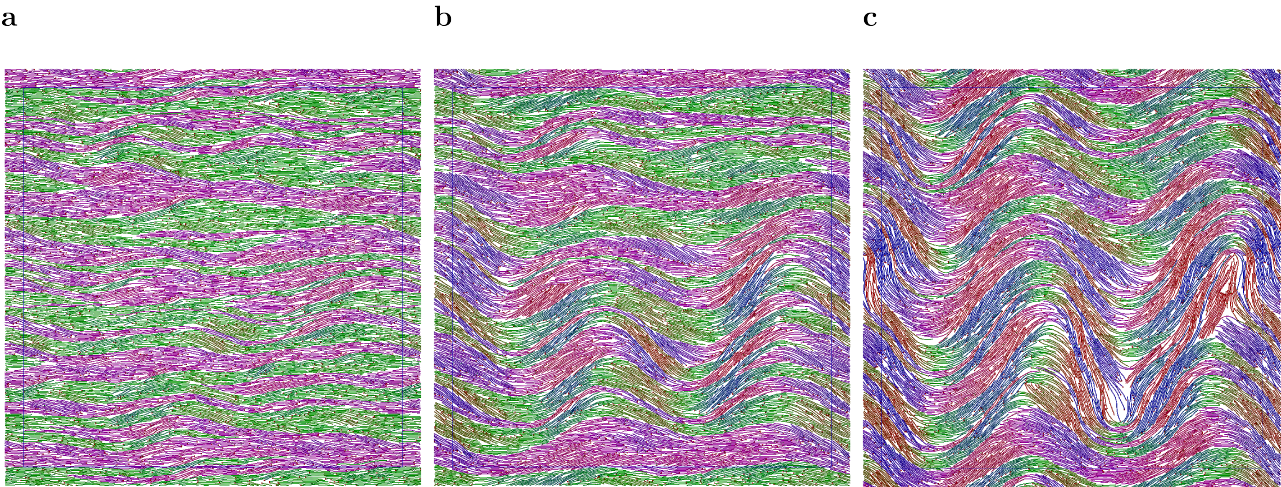}\\
\includegraphics[width=0.35\textwidth]{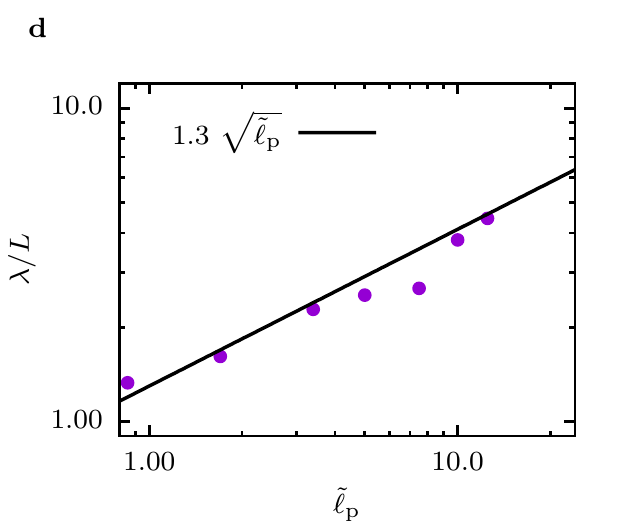}
\includegraphics[width=0.35\textwidth]{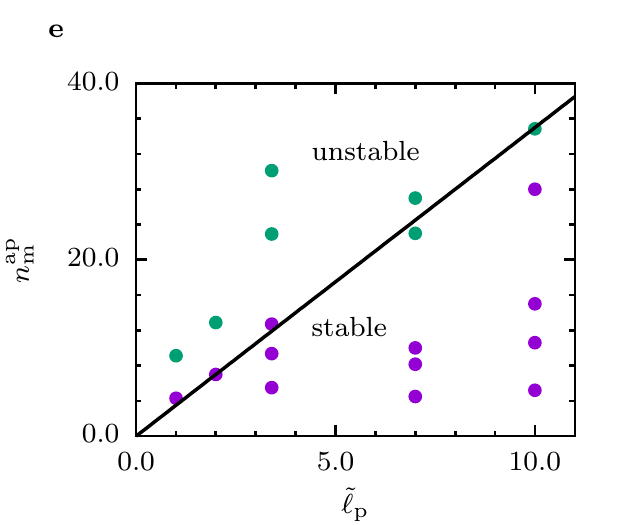}
\caption{{\bf Onset of domain formation: Euler-like buckling instability of polar bands.\\\hspace{\textwidth}} \textbf{a-c} Snapshots of the time evolution of a initially nematic system with $(\phi,\tilde{\ell}_\mathrm{p},\tilde{L})=(0.66,10,40)$, 
for $t/\tau_\mathrm{R}^0=0.1, 0.2, 0.3$ respectively. 
\textbf{d} Wavelength of the instability of polar-sorted bands as a function for varying persistence length and $(\phi,\tilde{L})=(0.66,40)$. 
\textbf{e} Stability phase diagram for various active forces and persistence lengths with paramaters $(\phi,\tilde{L})=(0.66,20)$.
The solid line separates different phases and indicates the values of the buckling force.}
\label{fig:instab}
\end{figure*}

The strength of the active force does not only determine the average filament velocity, 
but also structure, size, and stability of the domains. 
Polar bands are stable for weak active forces and large persistence lengths, whereas they become unstable and buckle at a particular 
wavelength $\lambda$ for sufficiently large active forces. 
The time evolution of the motor-filament mixture in Fig.~\ref{fig:instab}\textbf{a}-\textbf{c} shows polarity-sorted bands,
progressive bending and breaking of bands, see also Video~3.
The wavelength $\lambda$ at the instability displays a square-root dependence on the filament persistence length $\ell_\mathrm{p}$ (see fig.~\ref{fig:instab}\textbf{d}). 
          The assumption that the active force in all these systems evolves independent of the persistence length (see Fig.~SI~4) suggests an Euler buckling-type instability with buckling force $F_\mathrm{b} \propto E/\lambda^2$. Here $E\propto \ell_\mathrm{p}$ is the (effective) 
elastic modulus (Frank constant) \cite{ODIJK86}.  Together with $F_\mathrm{b} \sim {n_\mathrm{m}^\mathrm{ap}}$ this provides the scaling 
$\lambda^{2} \propto \ell_\mathrm{p}/{n_\mathrm{m}^\mathrm{ap}}$. 
This scaling was also found in wet active nematics ~\cite{RAMASWAMY010,THAMPI013,THAMPI014b,GIOMI015,HEMINGWAY016,GUILLAMAT017} 
with the same argument of balancing active stress and nematic elasticity. The investigation of the stability of the system with simulations 
at various combinations of  $\tilde{n}_\mathrm{m}$ and $\tilde{\ell}_\mathrm{p}$  allows us to determine the  phase diagram shown in Fig.~\ref{fig:instab}\textbf{e}. 
Here the stability limit of the polarity-sorted bands increases with increasing persistence length, 
which nicely agrees with the predicted linear dependence of the buckling force on $\tilde{\ell}_\mathrm{p}$.

\subsection*{Intradomain dynamics}
\begin{figure*}
\centering
\hspace*{0.6cm}\raisebox{0.18\height}{\includegraphics[width=0.23\textwidth]{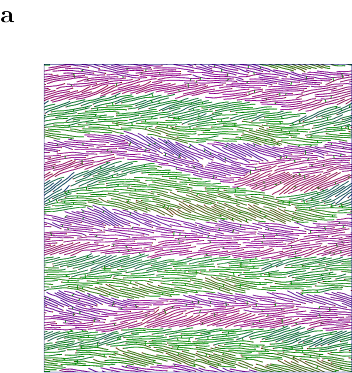}}
\hspace*{0.8cm}\includegraphics[width=0.35\textwidth]{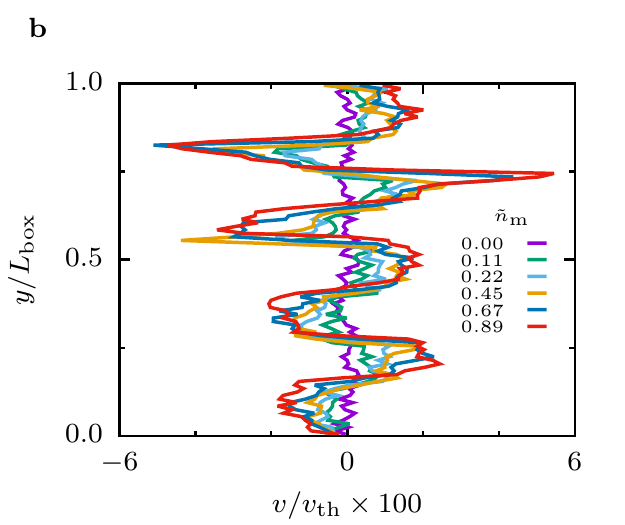}\\
\hspace*{0.3cm}\includegraphics[width=0.35\textwidth]{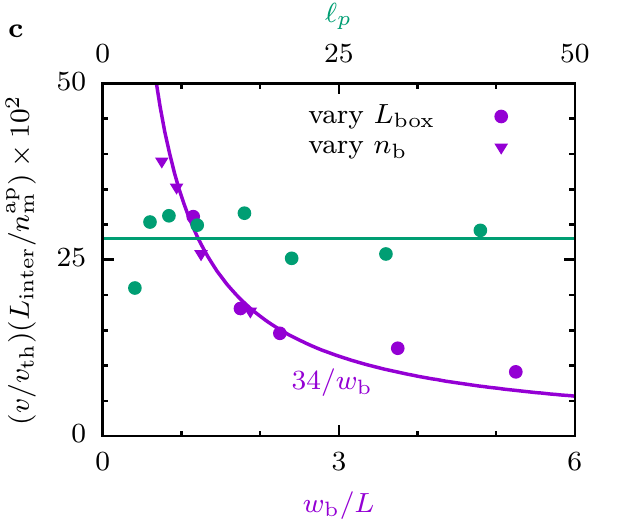}
\hspace*{-0.0cm}\includegraphics[width=0.35\textwidth]{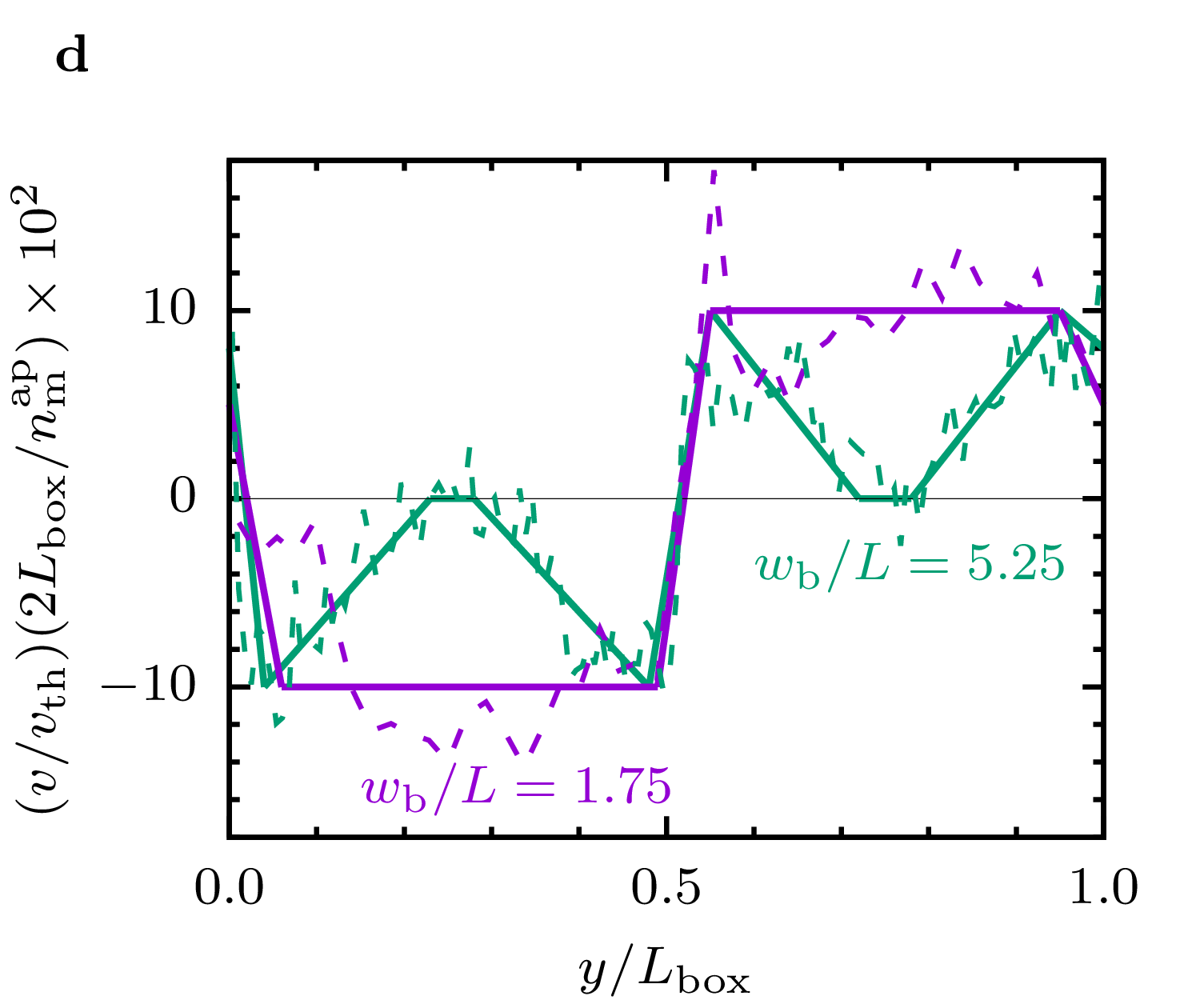} 
\caption{{\bf Intra-band filament dynamics for narrow and wide bands at steady state.\\\hspace{\textwidth}} 
\textbf{a} Snapshots for a system with parallel bands for $(\phi,\tilde{\ell}_\mathrm{p},n_\mathrm{b},\tilde{n}_\mathrm{m})=(0.66,20,8,0.12)$. 
\textbf{b} Velocity profiles for various $\tilde{n}_\mathrm{m}$. $(\phi,\tilde{\ell}_\mathrm{p},n_\mathrm{b})=(0.66,20,8)$ 
\textbf{c} Normalised averaged band velocities as a function of the band width $w_\mathrm{b}$ and 
the persistence length $\ell_\mathrm{p}$ of the filaments for $(n_\mathrm{b},\tilde{n}_\mathrm{m})=(6,0.45)$. The antiparallel motor density 
is ${n_\mathrm{m}^\mathrm{ap}}/L_\mathrm{inter}$ where $L_\mathrm{inter}=n_\mathrm{b} L_\mathrm{b}$ is the total length of the interface in the system.
\textbf{d} Normalised velocity profiles for two band widths. Dashed lines are simulation results and solid lines are guides 
to the eye.}
\label{fig:band_vel}
\end{figure*}

Our component-based model also provides detailed microscopic information about the filament dynamics 
within the polar domains.  We studied configurations
with a stable number of bands, an example of which is shown in Fig.~\ref{fig:band_vel}\textbf{a}. 
The corresponding velocity profiles are calculated for various values of the active force
by changing $n_\mathrm{m}$, see Fig.~\ref{fig:band_vel}\textbf{b}. The average band velocity increases linearly 
with ${n_\mathrm{m}^\mathrm{ap}}$, in agreement with the results in Fig.~\ref{fig:snapshots}\textbf{f}. 
The total length of the interface, $L_\mathrm{inter}\simeq n_\mathrm{b} \mathrm{b} L_\mathrm{box}$, varies by changing the number of bands, or by enlarging the box size. 
The band velocity is roughly independent of the persistence length, 
see Fig.~\ref{fig:band_vel}\textbf{c}, which indicates that the interfacial structure is not dramatically 
affected by the filament flexibility. 

The results shown above demonstrate that {the force applied by the antiparallel motors at the 
interfaces generates the motion of the bands}.  The parallel motors 
connecting filaments within the bands transmit this force in the perpendicular direction, but the friction with the
embedding medium strongly reduces the velocity in the center of the bands for wide bands. From velocity profile for the wide bands, see Fig.\ref{fig:band_vel}~\textbf{d}, we estimate the velocity decay length to be between half and two filament lengths. This is confirmed by an explicit  calculation of the velocity correlation length from the velocity correlation function as outlined in  fig.~SI~5, fig.~SI~6 and fig.~SI~7. Moreover, the velocity correlation length is rather insensitive to a change in activity as was also found in Ref. \cite{THAMPI013}, but does depend on the persistence length, see fig.~SI~8. 

For narrow bands, the velocity profiles 
of the two interfaces overlap, resulting in plug flow-type velocity profiles, as shown in Fig.~\ref{fig:band_vel}\textbf{d}.
Note that in most cases  the orientation of the filaments at the interface has a well-defined inclination 
angle ($15^\circ-25^\circ$), see for example Fig.~\ref{fig:band_vel}\textbf{a} or  Fig.~\ref{fig:snapshots}\textbf{c}. 
Thus motors between antiparallel filaments can only attach close to the filament ends imposing 
a pushing force from the rear end.  Closer to the center of a band, the filaments are oriented parallel to the interface. 


\subsection*{Dynamics of topological defects}
\begin{figure*}
\centering
\includegraphics[width=0.7125\textwidth]{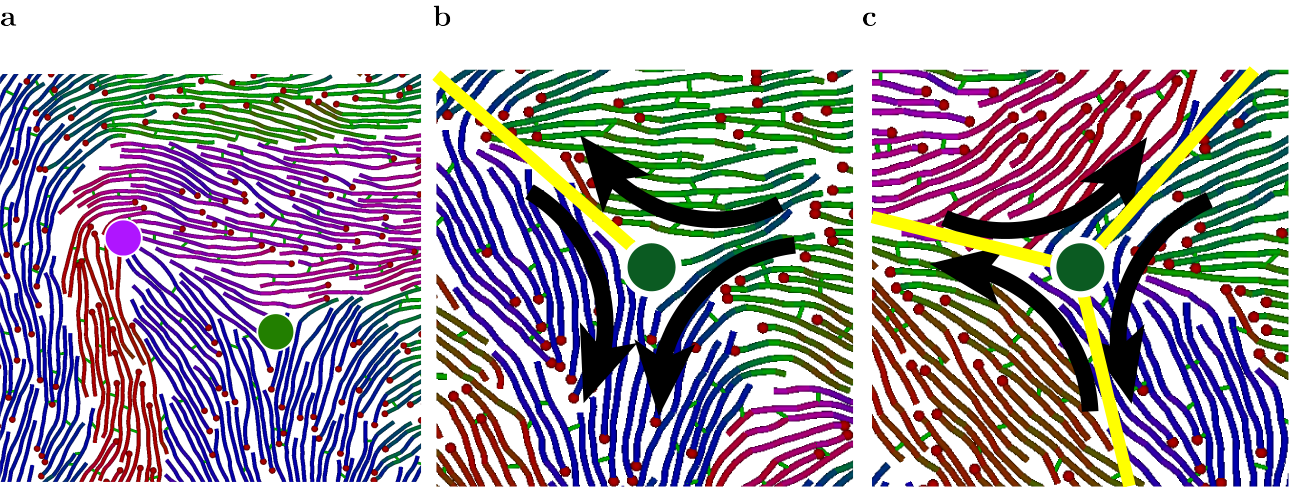}\\
\includegraphics[width=0.35\textwidth]{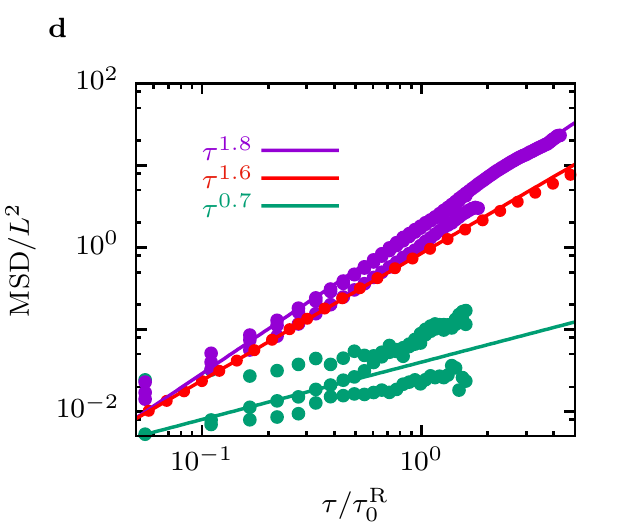}
\hspace*{-0.5cm}\includegraphics[width=0.35\textwidth]{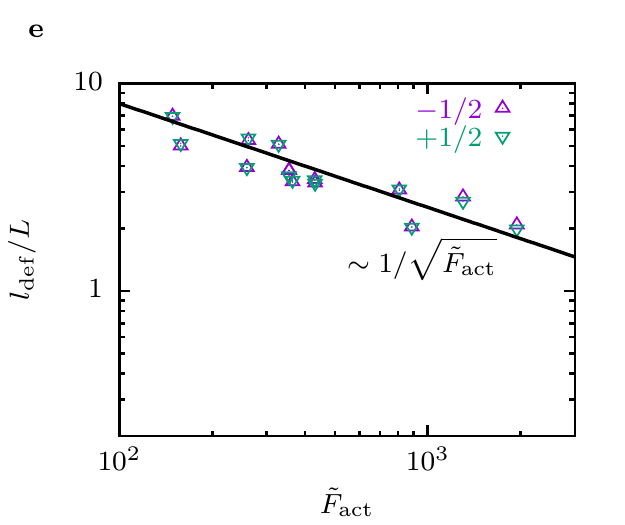}
\hspace*{-0.6cm}\includegraphics[width=0.35\textwidth]{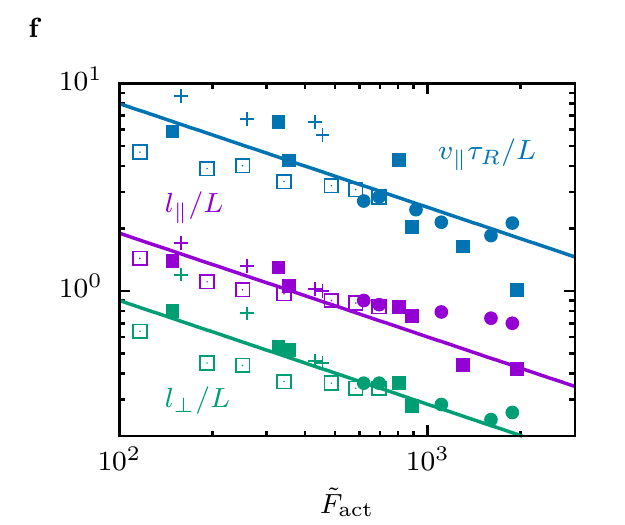}
\caption{{\bf Densities and mean squared displacements of topological +1/2 and -1/2 defects.\\\hspace{\textwidth}} \textbf{a} Snapshot of a pair of \nicefrac{-1}{2} (green) and a \nicefrac{+1}{2} (magenta) defects.
\textbf{b} Polar and nematic order near a \nicefrac{-1}{2} defect with $C_1$ symmetry. 
\textbf{c} Polar and nematic order near a \nicefrac{-1}{2} defect with $C_3$ symmetry. 
In (\textbf{a}), (\textbf{b}) and (\textbf{c}), filament colors indicate the polar angle of filament orientation.
In (\textbf{b}) and (\textbf{c}), the yellow lines indicate boundaries between polar domains. 
\textbf{d} MSD of the filament center-of-mass (red), \nicefrac{-1}{2} defects (green) and of  \nicefrac{+1}{2} defects (magenta). 
\textbf{e} Average distance between defects as a function of the active force.
\textbf{f} Estimation of the domain size with three different approaches. 
Solid lines show the $1/\sqrt{F_\mathrm{act}}$ dependence, and symbols simulations with data sets from the table in \protect{Fig.~\ref{fig:snapshots} 
with $(\phi,\tilde{L})=(0.66,20)$.}
}
 \label{fig:defects}
\end{figure*}

The system of polar filaments and motor proteins shares features with 
both active nematics, and polar active fluids, although it is clearly distinguishable from both of them.
As for active nematics, anti-parallel motion at the interfaces drives the dynamics; as for polar active fluids, polar order emerges
within the domains.
However the characteristic topological defects that appear in polar active fluids,  
$+1$ or $-1$ defects \cite{GIOMI014}, are  never observed here. 
Instead, the defect structures in the dynamic disordered  phase are 
$\nicefrac{+1}{2}$ and $\nicefrac{-1}{2}$ topological defects, as shown in Fig.~\ref{fig:defects}. 
A defect pair is formed when a polar band buckles by extensional forces, such that the convex side forms a 
\nicefrac{+1}{2} defect, while the concave side forms a \nicefrac{-1}{2} defect, see Fig.~\ref{fig:defects}\textbf{a}.
Importantly, these are not the standard defects of active nematics, because the three domains which meet at a
\nicefrac{-1}{2} defect display polar order, which implies that there can be active forces where domains with
anti-parallel polar order meet. 
In particular, our polar filament model gives rise to two types of \nicefrac{-1}{2} defects 
with different orientations of the polar domains around the defect core. 
One has $C_1$ symmetry, the other $C_3$ symmetry, see Figs.~\ref{fig:defects}\textbf{b} and \textbf{c}. 
Note that for the $C_1$-defect in Fig.~\ref{fig:defects}\textbf{b}, there are additional (unmarked) boundaries between anti-parallel 
domains in the lower left and right corners where the motion is originated. 


The topological defects are calculated following Refs.~\cite{BATES08,DECAMP015} and show super-diffusive motion of the \nicefrac{+1}{2} defects, and sub-diffusive 
motion of the \nicefrac{-1}{2} defects as follows from the defects  MSDs in Fig.~\ref{fig:defects}\textbf{d} and illustrated in Video~4. The annihilation of two defects is shown in Video~5.  
However, we do not find any differences in the dynamic behaviour of the two types of \nicefrac{-1}{2} defects.
The defect density depends linearly on the activity, which is reflected by the inverse square root 
dependence of the distance $l_\mathrm{def}$ between the defects on the force as shown in Fig.~\ref{fig:defects}\textbf{e}. 
A similar scaling is found for a related characteristic length scale in the system, the domain size, 
shown in Fig.~\ref{fig:defects}\textbf{f}.  
The exponential decay of the parallel and perpendicular spatial orientational correlation functions
$\widehat{\Omega}_\mathrm{\parallel}(r)$ and $\widehat{\Omega}_\mathrm{\perp}(r)$ (see fig.~SI~9) provides the 
correlation lengths $l_\perp$ and $l_ \parallel$, which are estimations for width and length of the domains.   
An independent estimate is $v_{\parallel}\tau_\mathrm{R}$ \cite{STARK016}, which is proportional to the distance that a filament 
moves along the domain boundary, before changing direction. 
Note that both $\tau_\mathrm{R}$ and $v_\parallel$ are averaged values over a distribution of faster (antiparallel) 
and slower (parallel) filaments.  Although these three lengths provide different quantitative estimations of the
domain size, their scaling is consistent with an inverse square-root dependence on the activity. 

\subsection*{Universality of domain-size scaling and active diffusion}
\begin{figure*}
\centering
\includegraphics[width=0.9\textwidth]{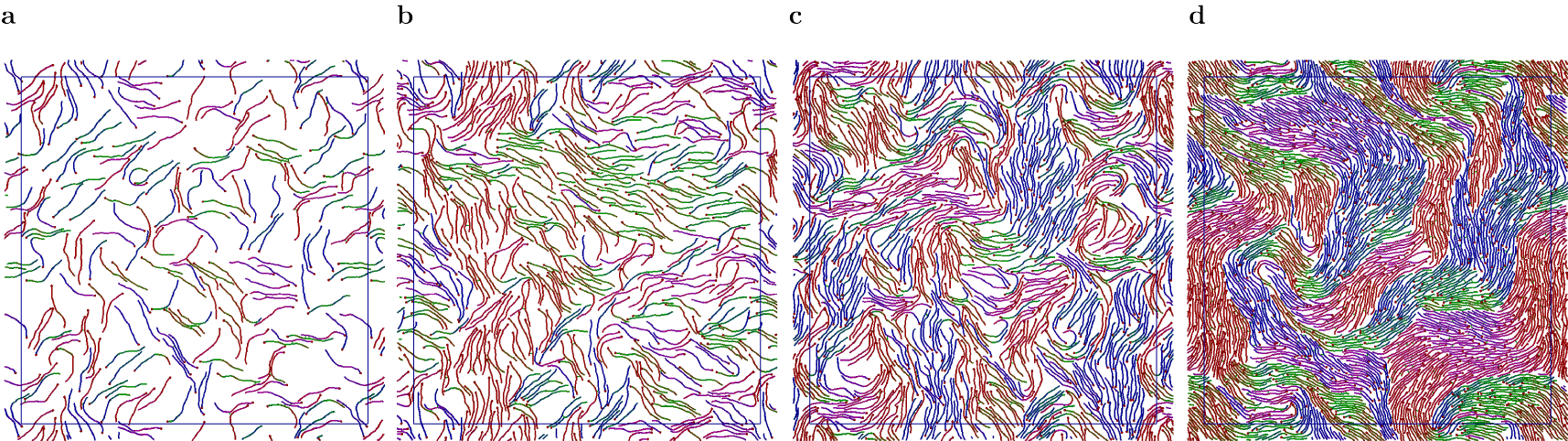}\\
\includegraphics[width=0.35\textwidth]{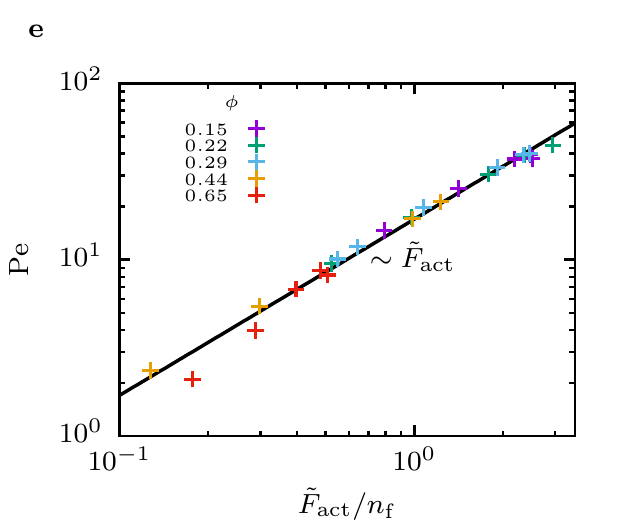}
\hspace*{-0.5cm}\includegraphics[width=0.35\textwidth]{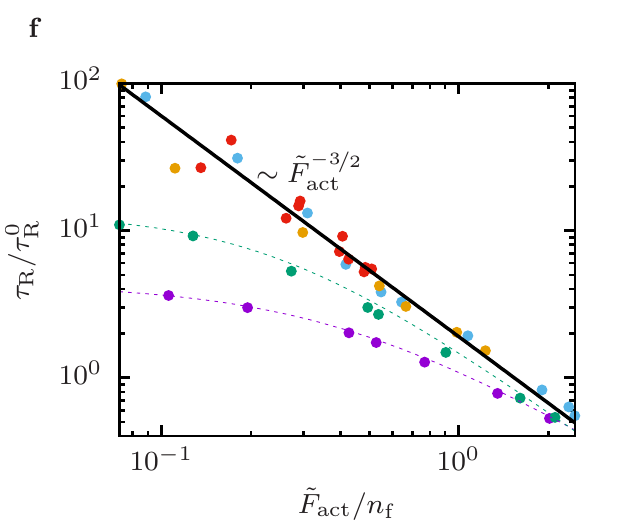}
\hspace*{-0.6cm}\includegraphics[width=0.35\textwidth]{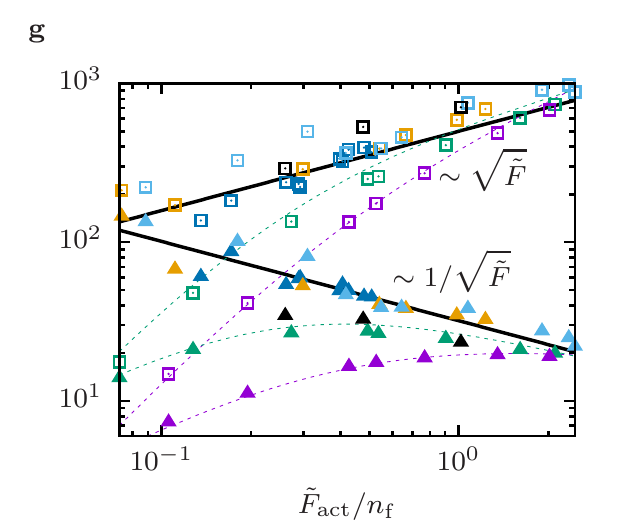}
\caption{{\bf Filament density-dependent structure formation and single-filament dynamics.\\\hspace{\textwidth}}
 \textbf{a}-\textbf{d} Snapshots for systems at various densities, $\phi=(0.15,0.29,0.44,0.66)$, 
where $n_\mathrm{m}$ is varied to result in similar $n_\mathrm{m}^\mathrm{ap}/n_\mathrm{f} \approx 0.165$. 
\textbf{e-g} Dimensionless observables as a function of the active force for different area fractions. 
Symbols correspond to simulation data, dashed lines are to guide the eye and full lines indicate limiting powerlaws at large densities: 
\textbf{e} Peclet number, 
\textbf{f} active rotation time,
\textbf{g} active diffusion coefficient $v_\parallel^2\tau_\mathrm{R}$ (open squares), and domain size $v_\parallel \tau_\mathrm{R}$ (solid up-triangles). 
Colored labels in (\textbf{e}) refer to densities in (\textbf{e}, \textbf{f}, \textbf{g}). 
Black symbols in (\textbf{g}) correspond to simulations with longer filaments with $(\phi,\tilde{\ell}_\mathrm{p},\tilde{L})=(40,0.66,1.7)$.}
\label{fig:density}
\end{figure*}

In passive lyotropic liquid crystalline systems particle density and shape determine both structural and dynamical properties. 
For active systems, however, the active force provides an additional control parameter.  
Figure ~\ref{fig:density}\textbf{a}-\textbf{d} shows simulation snapshots for various filament densities and different $\tilde{n}_\mathrm{m}$, 
chosen such that ${n_\mathrm{m}^\mathrm{ap}}/n_\mathrm{f}$ is comparable so that the systems have similar active force densities. We observe a gradual 
change from small local bundles held together by motors at low densities, to a dense disordered nematic dominated by 
excluded-volume interactions similarly as  in experiments \cite{GUILLAMAT017}. Although the structures observed are very different, the (parallel) velocities still show a unique 
linear scaling with the force density $v_{\parallel} \propto {\rm Pe} \propto {F_\mathrm{act}/n_\mathrm{f}}$, see Fig.~\ref{fig:density}\textbf{e}.
However, the active rotation times shown in Fig.~\ref{fig:density}\textbf{f} show a significant dependence on the density, 
especially for small densities and low activities, where the active rotation time is similar to the passive one. 
Nevertheless, for large activities and for large densities again a unique scaling behaviour 
$\tau_\mathrm{R} \sim (\tilde{F}/n_\mathrm{f})^{\nicefrac{-3}{2}}$ is observed, see also Fig.~\ref{fig:D_act}\textbf{b}. 
Therefore, at large densities or large activities, the active diffusion coefficient $D_\mathrm{act}\sim v_{\parallel}^2 \tau_\mathrm{R}$ 
and domain size $v_{\parallel}\tau_\mathrm{R}$ show a universal scaling, see Fig.~\ref{fig:density}\textbf{g}. 
The chosen scaling variables ($F_\mathrm{act}/n_\mathrm{f}$, $\tau_\mathrm{R}/\tau_\mathrm{R}^0$ and $v_\parallel L/D_0$) are independent 
of the filament length, as indicated by of three simulations for filaments twice as long as all our other simulations.

\section*{Discussion}

Our simulations of semiflexible polar filaments crosslinked by 
molecular motors generate macroscopic active domains  and defect formation in 
the absence of hydrodynamic interactions. The simulations reveal a novel structure, an '{\it active polar nematic}', which consists of nematically arranged polar domains.
Essential results of the simulations are that activity manifests at domain boundaries, not inside domains, and that new types of topological defects appear,  which are  distinguished by the two possible arrangements of polar orientation at a three-fold junction. 
An initially nematic arrangement of filaments is unstable and is found to evolve via polarity sorting, band formation, and buckling into a stationary but highly dynamic polar-domain structure with persistent defect formation and annihilation. 
We find universal scaling of active filament diffusion and of domain sizes with the active force determined by the number of antiparallel motors and their extension. 
Our model can readily be extended to three-dimensional systems, to study the effects of stiff and (semi)flexible 
confinements, as well as of models with an increased level of complexity like the inclusion of passive crosslinkers 
or hydrodynamic interactions.


\section*{Materials and Methods}
{Suspensions of $n_\mathrm{f}$ semiflexible filaments and $n_\mathrm{m}$ motors are studied using Langevin dynamics \cite{JENSEN013} in two dimensions with periodic boundary conditions. Semiflexible filaments of length $L$ are modelled as discrete chains of $n_\mathrm{s}$ beads of mass $m$ with position vectors ${\mathbf r}_q^i$ where $q=\{1....n_\mathrm{f} \}$ and $i=\{1....n_\mathrm{s} \}$.  
The $n_\mathrm{s}$ beads define $n_\mathrm{s}-1$ segments of length $a_0=L/(n_\mathrm{s}-1)$ with unit tangent vectors $\hat{{\mathbf t}}_q^i=(\mathbf{r}_q^{i+1}-\mathbf{r}_q^{i})/|\mathbf{r}_q^{i+1}-\mathbf{r}_q^{i}|$.   
The beads are connected by harmonic bonds with equilibrium length $a_0$ and spring constant $k r_\mathrm{0}^2/\mathrm{k}_\mathrm{B} T=250$. A bending potential $V_\mathrm{b}=\kappa(1-\hat{{\mathbf t}}_q^i \cdot \hat{{\mathbf t}}_q^{i+1} )/a_0$ gives rise to a tangent vector correlation function $\left<{\bf t}(0)\cdot{\bf t}(s)\right >=\exp[-s/\ell_\mathrm{p}]=\exp[-(d-1)\mathrm{k}_\mathrm{B} T s/2\kappa]$ with dimensionality $d$ and arclength $s$ along the contour, such that the bare persistence length is $\ell_\mathrm{p}/L =2\kappa/L\mathrm{k}_\mathrm{B} T$ \cite{KIERFELD04}. Non-bonded inter and intramolecular excluded volume and attractive interactions are described by a Lennard Jones potential $V_{LJ}=\varepsilon[(\sigma/r)^{12}-2(\sigma/r)^6]$ where $r$ denotes the inter or intra molecular bead-bead separation, $\varepsilon$ the interaction strength, $\sigma$ the interaction diameter, i.e. the beads are not overlapping) and  $r_c=3\sigma$ the potential cut-off radius. In all our simulations $k_\mathrm{B}T/\varepsilon=10$, i.e. attractions between filaments are negligible. 
These parameters lead to an effective Barker-Henderson diameter \cite{BARKER76} which is 14\% less than that for a system with a purely repulsive WCA potential at $\varepsilon/k_\mathrm{B}T=1.0$ and 4\% less than the WCA potential with  $\varepsilon/k_\mathrm{B}T=0.1$.
The filament beads also act as binding sites for 
molecular motors. These are modelled as harmonic springs $V_\mathrm{m}=k_\mathrm{m}(r-r_\mathrm{0})^2/2$ with spring constant $k_\mathrm{m}$ and equilibrium length $r_\mathrm{0}=\sigma$ that walk on the filaments \cite{HEAD011,GAO015,BETTERTON016,RAVICHANDRAN017}. Note that here the discretisation length $a_0$ directly sets the motor step size. 
When the separation between two empty sites on nearby filaments is smaller than $r_\mathrm{att}$  a free (unbound) motor can bind at a rate $(p_\mathrm{att} /\delta t) \ (n_\mathrm{m}^\mathrm{f}/n_\mathrm{m})$ where $n_\mathrm{m}^\mathrm{f}$ is the number of free motors and $p_\mathrm{att}=1$ in all simulations. When bound, the $n_\mathrm{m}^\mathrm{b}$ motors walk in the direction of filament polarity (from segment $1\rightarrow n_\mathrm{s}$) in discrete steps of size $a_0$ at a rate $\Gamma= p_\mathrm{m}^0/\Delta t \ \exp{(-\beta \Delta E)}$. Here $p_\mathrm{m}^0$ is the probability for an attempted step of a  motor attached to two filaments, $\Delta t$ the motor time-step and $\Delta E$ the energy difference before and after the motor has stepped.  Motors detach only when one leg reaches the end of the filament or when the motor is  stretched beyond a length $r_\mathrm{off}=3.25\ r_0$. After detaching, the free motors form a bath with homogeneous concentration $n_\mathrm{m}^\mathrm{f}/L_\mathrm{box}^2$ where $L_\mathrm{box}$ is the box size. The Langevin equation of motion for each bead $i$ on filament $q$,
\begin{equation}
m\frac{\mathrm{d}^2 {\bf r}_q^i}{\mathrm{d} t^2} = {\bf F}_q^i - \gamma \frac{\mathrm{d} {\bf r}_q^i}{\mathrm{d} t} + {\bf R}_q^i,\label{eq:lang}
\end{equation}
is integrated with a time step $\delta t$ using the integrator proposed in Ref.~\cite{JENSEN013}. In order to satisfy the fluctuation-dissipation theorem the random forces are Gaussian distributed with mean zero and variance ${2\mathrm{k}_\mathrm{B} T \gamma \delta(t-t')}$. The integration time step $\delta t$ and friction $\gamma$, were chosen such that $\delta t/m = 0.005$ and $\gamma \delta t/m \ge 0.005$. For these parameters the (passive) center-of-mass motion of a single filament is diffusive for center-of-mass displacements larger than a fraction of the length of a filament,  i.e., for center-of-mass diffusion the dynamics is essentially overdamped. Note that the simulation time increases linearly with the friction constant.  Simulations typically contain 900 filaments of 21 beads and up to 3200 motors.

The coupling of filaments by molecular motors leads to sliding and binding forces that depend on the relative orientation of the filaments, see Fig.~\ref{fig:snapshots}\textbf{a},\textbf{b}, where antiparallel motors (coupling two antiparallel filaments) exert larger forces than parallel motors (coupling two parallel filaments). 
Moreover, the activity induced by dimeric motors (one motorarm is grafted) tetrameric motors (both motor arms move simultaneously)  is larger \cite{RAVICHANDRAN017}.  The motor model here is a hybrid of the two, both motor arms can move with equal probability but only one at a time. 
 
Rotational diffusion is measured through the filament orientation correlation function, where the orientation is defined as the eigenvector corresponding to the largest eigenvalue of the moment of inertia tensor.

\section*{Supplementary Materials}

\noindent{Movie S1. Dynamics of polarity sorting and coarsening of polar bands from the initial disordered nematic state.}\\

\noindent{Movie S2. Steady state dynamics of polar domains with repeated creation and annihilation of defect pairs.}\\

\noindent{Movie S3. Dynamics of the  elastic instability.  Formation and buckling of  polar bands leads to the emergence of the disordered phase.}\\

\noindent{Movie S4. Defect dynamics. Extensile motion of a $\pm$\nicefrac{1}{2} defect pair.}\\

\noindent{Movie S5. Annihilation of a \nicefrac{+1}{2} (blue) and \nicefrac{-1}{2} defect (purple).}\\

\noindent{Fig. S1. Time sequence of filament self-organisation.}\\

\noindent{Fig. S2. Normalised  parallel velocity $v_\parallel/v_\mathrm{th}$.}\\

\noindent{Fig. S3. Time evolution of the number of antiparallel motors.}\\

\noindent{Fig. S4. The number of antiparallel motors as a function of  the total number of motors.}\\

\noindent{Fig. S5. Segment orientational correlation functions.}\\

\noindent{Fig. S6. Filament velocity correlation functions for three motor densities.}\\

\noindent{Fig. S7. Time scale renormalization for active motion.}\\

\noindent{Fig. S8. Amplitude and velocity decorrelation length of the velocity correlation function
  function.}\\

\noindent{Fig. S9. Dependence of the velocity correlation length on the persistence length.}\\

\bibliography{filmot}
\bibliographystyle{ScienceAdvances}

\noindent \textbf{Acknowledgements:} 
%
The authors thank Roland G Winkler for helpful conversations.\\
\noindent \textbf{Author Contributions:} GV, TA and GG conceived the research. GV wrote all simulation codes, performed simulations and analysed the data. All authors wrote the manuscript.\\
\noindent \textbf{Competing Interests:} The authors declare that they have no competing financial interests.\\
\noindent \textbf{Data and materials availability:} Additional data and materials are available online.


\end{document}


\title{Filamentous Active Matter: Band Formation, Bending, Buckling, and Defects\\
Supplementary Information} \author{G. A. Vliegenthart, A. Ravichandran, M. Ripoll, T. Auth,  G. Gompper}
\affiliation{Theoretical Soft-Matter and Biophysics, Institute for Advanced Simulations, 
Institute of Complex Systems, \\
Forschungszentrum J\"ulich, 52425 J\"ulich, Germany}

\maketitle

\noindent{\bf Video captions}

\vspace*{-0.5cm}

\begin{video}[h]
\caption{{\bf Dynamics of polarity sorting and coarsening of polar bands from the initial disordered nematic state.\\\hspace{\columnwidth}} Here $(\phi,\ellpt,\nmt)=(0.66,5,0.89)$. The duration of the movie is $\tilde{t}\approx 2.7 \tau_\mathrm{R}^0$ (500 frames, sampling every 200 time steps).}
\label{mov:polar_sorting}
\end{video}

\vspace*{-1cm}

\begin{video}[h]
\caption{{\bf Steady state dynamics of polar domains with repeated creation and annihilation of defect pairs.\\\hspace{\columnwidth}}  Here $(\phi,\ellpt,\nmt)=(0.66,3.4,0.89)$.  The duration of the movie is $\tilde{t} \approx 13.5 \tau_\mathrm{R}^0$ (500 frames, sampling every 1000 time steps).}
\label{mov:disordered}
\end{video}

\vspace*{-1cm}

\begin{video}[h]
\caption{{\bf Dynamics of the  elastic instability.\\\hspace{\columnwidth}} Formation and buckling of  polar bands leads to the emergence of the disordered phase. Here the parameters are $(\phi,L/\sigma,\ellpt,\nmt)=(0.66,40,10,1.0)$. The duration of the movie is $\tilde{t} \approx 2 \tau_\mathrm{R}^0$ (800 frames, sampling every 500 time steps).}
\label{mov:buckling}
\end{video}

\vspace*{-1cm}

\begin{video}[h]
\caption{{\bf Defect dynamics.\\\hspace{\columnwidth}} Extensile motion of a $\pm$\nicefrac{1}{2} defect pair. The \nicefrac{-1}{2} defect (purple) is nearly immobile, the \nicefrac{+1}{2} defect (blue) moves active diffusive.  Here the parameters are $(\phi,\ellpt,\nmt)=(0.66,5,0.89)$. The duration of the movie $\tilde{t} \approx 2.2 \tau_\mathrm{R}^0$. This movie is clipped from a larger field of view (70 frames, sampling every 2000 time steps).}
\label{mov:defects_ext}
\end{video}

\begin{video}[h]
\caption{{\bf Defect annihilation.\\\hspace{\columnwidth}}Annihilation of a \nicefrac{+1}{2} (blue) and \nicefrac{-1}{2} defect (purple).  The parameters are $(\phi,\ellpt,\nmt)=(0.66,5,0.89)$. 
The duration of the movie is $\tilde{t} \approx 2.2 \tau_\mathrm{R}^0$. This movie is clipped from a larger field of view (70 frames, sampling every 2000 time steps).}
\label{mov:defects_anni}
\end{video}

\begin{figure}[H]
\begin{center}
\includegraphics[width=0.298\textwidth]{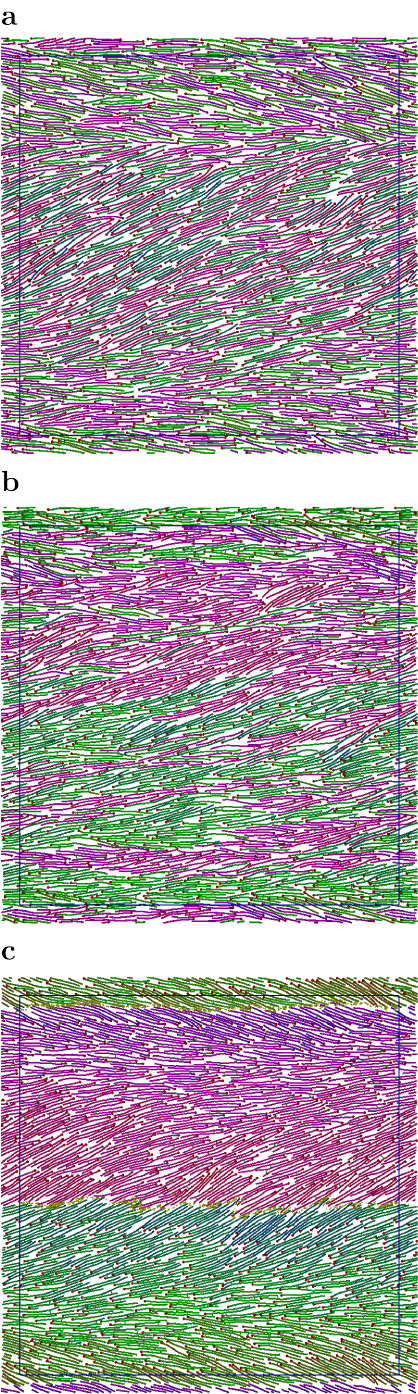}
\caption{{\bf Time sequence of filament self-organisation.\\\hspace{\columnwidth}} Here,
  $(\phi,\ellpt,\tilde{n}_\mathrm{m})=(0.66,20,0.89)$, with time
  $\tilde{t}=t D_0/L^2$ and $D_0$ the passive single filament diffusion
  coefficient. (\textbf{a})~$\tilde{t}=0$, nematic phase at zero
  activity; (\textbf{b})~$\tilde{t}=0.0125$ polarity sorting; 
  (\textbf{c})~$\tilde{t}=2.5$ stationary state consisting of
  stable bands.}
\label{fig:snapshots_SI}
\end{center}
\end{figure}

\noindent{\bf Band formation and steady state structure at large persistence lengths  or  low activity}\\
Figure~\ref{fig:snapshots_SI} shows the time evolution of a nematic
phase with large persistence length when motors are added.
At short times (\textbf{a}) the system is in a disordered nematic state. 
At intermediate times (\textbf{b}) the system forms polar sorted bands 
 that subsequently coarsen, and at late times (\textbf{c}), when the system reaches steady state, 
two stable oppositely aligned bands form. This is a typical time sequence for large persistence lengths and/or for small activities.

\begin{figure}[t!]
\includegraphics[width=0.4\textwidth]{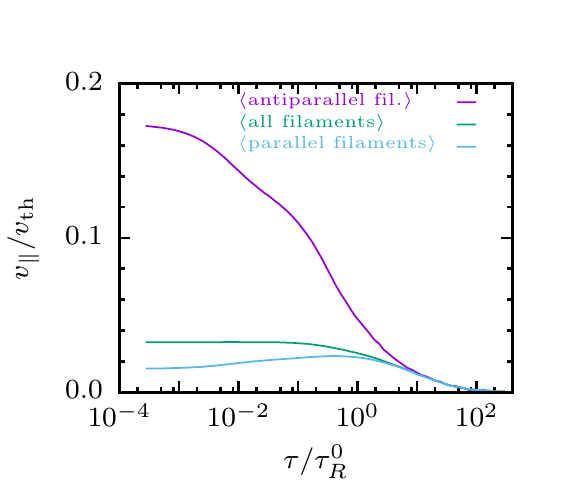}
\caption{{\bf Normalised  parallel velocity $\bf v_\parallel/v_\mathrm{th}$.\\\hspace{\columnwidth}}
Parallel velocity as a of function lag time $\tau/\tau_\mathrm{R}^0$ for $(\phi,\ellpt,\nmt)=(0.66,3.4,0.89)$. Here $v_\mathrm{th}=\sqrt{k_\mathrm{B}T/m}$.  Averages are shown for all filaments, for parallel filaments only and for antiparallel filaments only.}
\label{fig:vparallel}
\end{figure}

\newpage
\noindent{\bf Parallel velocity and temporal  orientational autocorrelation function}\\
Figure~\ref{fig:vparallel}\textbf{a} shows the parallel
velocity (averaged over all filaments) (green) as a function of the
lag time $\tau$. The average is also performed over only the filaments in a parallel environment (filaments point in the same direction)
(blue), and in a antiparallel environment (filaments at the interface point in opposite direction) (purple) filaments.  Clearly, the
velocity of the antiparallel filaments is much larger than that of the
parallel ones. The average velocity over all filaments is closer
to the average over the parallel filaments since the fraction of
parallel filaments is much larger. In each domain, parallel filaments
form the bulk of the domain interior, while antiparallel filaments are
found at the boundaries.  
\\

\begin{figure}[t!]
\includegraphics[width=0.4\textwidth]{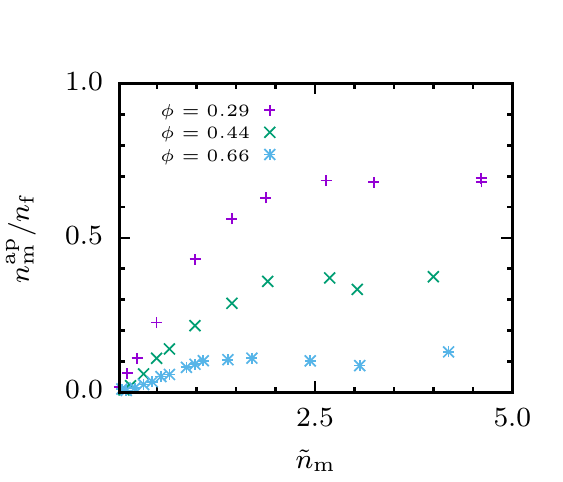}
\caption{{\bf The number of antiparallel motors as a function of the total number of motors.\\\hspace{\columnwidth}} Here ($\phi$, $\ellpt$, $L/\sigma$)=(0.66, 3.4, 20). }
\label{fig:nm}
\end{figure}

\noindent{\bf Fraction of antiparallel motors}\\
Figure~\ref{fig:nm} shows the number $\nmap$ of antiparallel motors 
per filament as a function of the total number of motors  $\nm$ in the system
for three filament densities.  In all cases the number of antiparallel
motors increases linearly with $\nm$ and saturates at larger $\nm$
where motors collide and detach from the filaments. 
\\

\begin{figure}[t!]
  \includegraphics[width=0.4\textwidth]{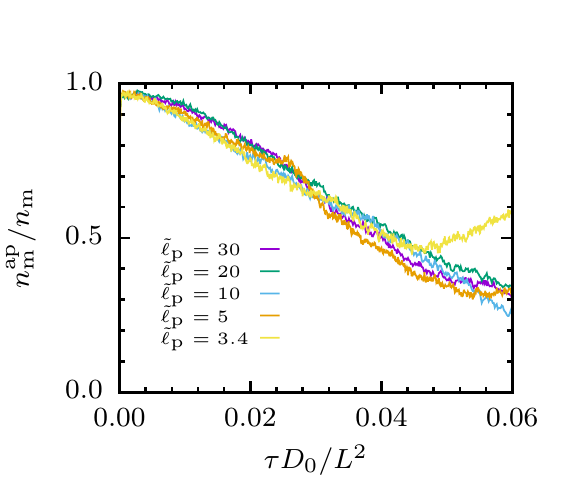}
  \caption{{\bf Time evolution of the number of antiparallel motors for different $\ellpt$.\\\hspace{\columnwidth}}Here ($\phi,\nmt,L/\sigma$)=(0.66,1.0,40). $\tau=0$ corresponds to the initial disordered nematic state.}
\label{fig:force_time}
\end{figure}

For systems evolving from the initial nematic state towards the
buckling of bands, the number of antiparallel motors varies non-monotonically 
 as a function of time for various
persistence lengths. For very short times, 
the parallel and antiparallel orientations of neighbouring
filaments  are equiprobable and the fraction of antiparallel
motors is $\nmap/\nm=1/2$. Rapidly, thin bands (width a few filament
thicknesses) appear and $\nmap/\nm$ shoots up to almost unity (not shown) after
which coarsening of the bands leads to a steady decrease of $\nmap$ with time as is shown in Fig.~\ref{fig:force_time}.
As the active force is proportional to $\nmap$,
Fig.~\ref{fig:force_time} also shows the time evolution of the force
on the filaments. Interestingly, the curves for different persistence
lengths follow the same time evolution, indicating that, at this stage,
the microstructures are very similar.
\\

\begin{figure}[t!]
\includegraphics[width=0.4\textwidth]{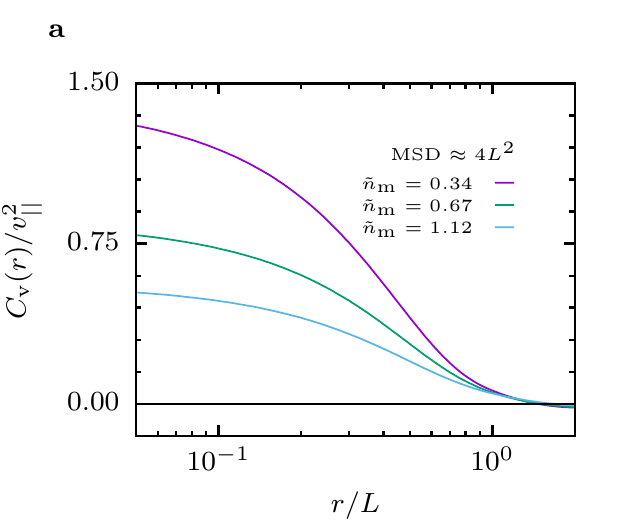}\\
\includegraphics[width=0.4\textwidth]{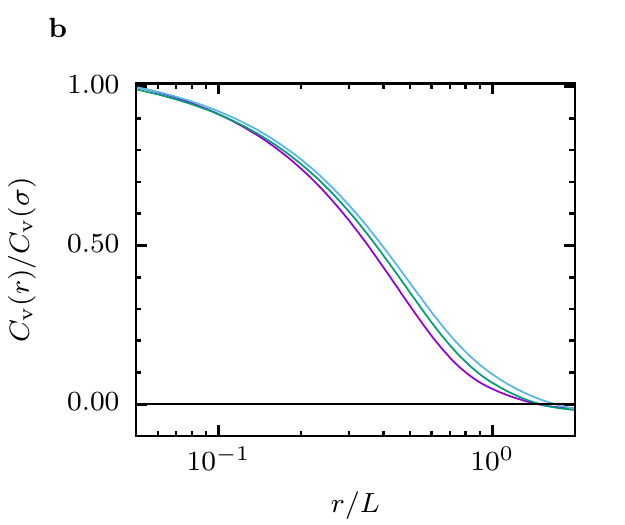}
\caption{{\bf Filament velocity correlation functions.\\\hspace{\columnwidth}}Velocity correlation functions for three motor densities at
  $\tau/\tau_\mathrm{R}^0\approx 4$ with ($\phi$, $\ellpt$)=(0.66,
  3.4). (\textbf{a}) Unnormalized correlation functions, 
nondimensionalised by the parallel velocity $v_{\parallel}^2$. (\textbf{b}) Normalised correlation functions. }
\label{fig:dcfnm}
\end{figure}

\begin{figure}[t!]
\includegraphics[width=0.4\textwidth]{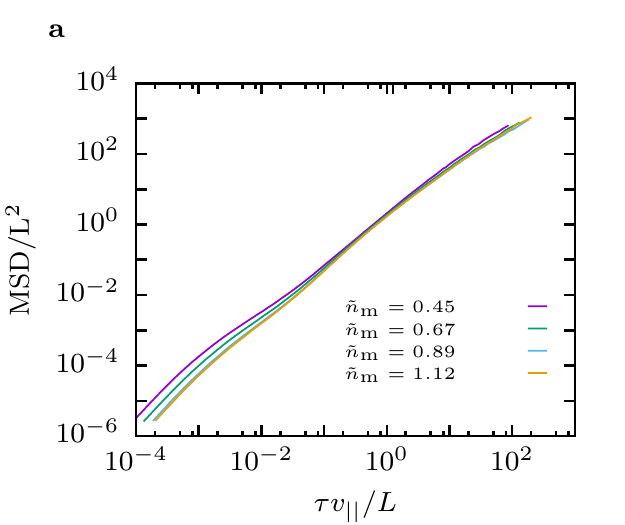}\\
\includegraphics[width=0.4\textwidth]{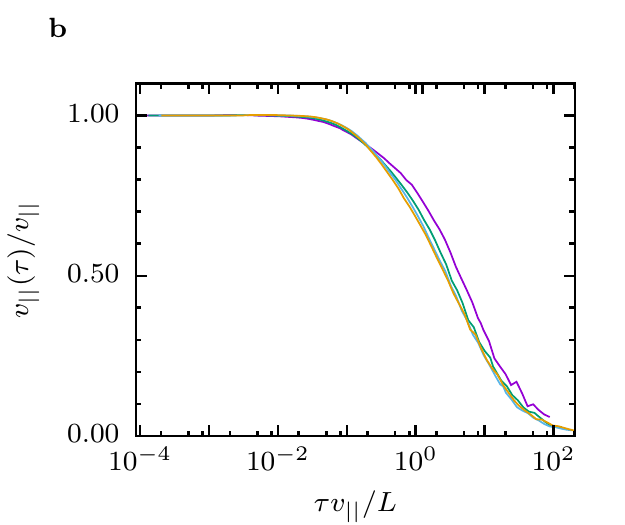}
\caption{{\bf Time scale renormalization for active motion.\\\hspace{\columnwidth}} \textbf{a}) MSD for different motor concentrations and (\textbf{b}) scaled parallel velocity for different motor concentrations. In both cases ($\phi$, $\ellpt$)=(0.66, 3.4). Note that $v_{\parallel}(\tau)$ refers to the time-dependent quantity and $v_{\parallel}=v_{\parallel}(\tau=0)$.}
\label{fig:dcf_msd}
\end{figure}

\noindent{\bf Velocity correlation function}\\
Additional comparison with existing experimental and simulation
work~\cite{SANCHEZ012,THAMPI013} can be done through the
spatial velocity correlation function
\begin{equation}
C_\mathrm{v}(r,\tau)=\frac{\av{\sum_{i,j\neq i} {\bf d}_i(\tau) \cdot {\bf d}_j(\tau) \delta(r-|{\bf r}_i-{\bf r}_j|)}_t} {\tau^2 \av{\sum_{i,j\neq i}\delta(r-|{\bf r}_i-{\bf r}_j|)}_t}. 
\end{equation}
 The center of mass displacement ${\bf d}_i$ of filament $i$ over a lag time $\tau$ is
defined as ${\bf d}_i(\tau)={\bf r}_i(t+\tau)-{\bf r}_i(t)$ with ${\bf
  r}_i(t)$ the center of mass of filament $i$ at time $t$.  
The velocity ${\bf v}={\bf d}(\tau)/\tau$ depends on the lag time $\tau$. 
Examples of $C_\mathrm{v}(r,\tau)$ for different activities at fixed
lag time are shown in Fig.~\ref{fig:dcfnm}\textbf{a}.  
The normalised velocity correlation function at a particular lag time in
Fig.~\ref{fig:dcfnm}\textbf{b}) is essentially independent of the
activity, which is consistent with previous experimental and numerical studies~\cite{SANCHEZ012,THAMPI013}.
The lag time $\tau$ has to be rescaled to compare $\mathrm{MSD}$, $v_{\parallel}(\tau)$ and correlation functions for different parameter sets.
The rescaled time evolution displayed in Fig.~\ref{fig:dcf_msd} suggests that $L/v_{\parallel}$ is the natural intrinsic time scale so that $\tau v_{\parallel}/L$ is the appropriate scaling variable
for the active ballistic regime where $\mathrm{MSD}\propto v_{\parallel}^2\tau^2$.

To gain more understanding, we focus on $r>\sigma$ and lag times $\tau$,
such that the passive system is in the diffusive regime (in
Fig.~\ref{fig:dcf_msd}, this corresponds to the time window in which
the scaled MSDs overlap).  In this case, long-ranged spatial
correlations build up \cite{WYSOCKI014} and $C_\mathrm{v}(r,\tau)$ typically shows an
exponential decay~\cite{DOLIWA00}, $C_\mathrm{v}(r,\tau)=A\exp[-r/\xi]$. 
Amplitudes $A$ and decay lengths $\xi$ for various parameter sets reveal the
dependence of the collective filament motion on motor densities, filament densities 
and persistence lengths. 

\begin{figure}[H]
\includegraphics[width=0.39\textwidth]{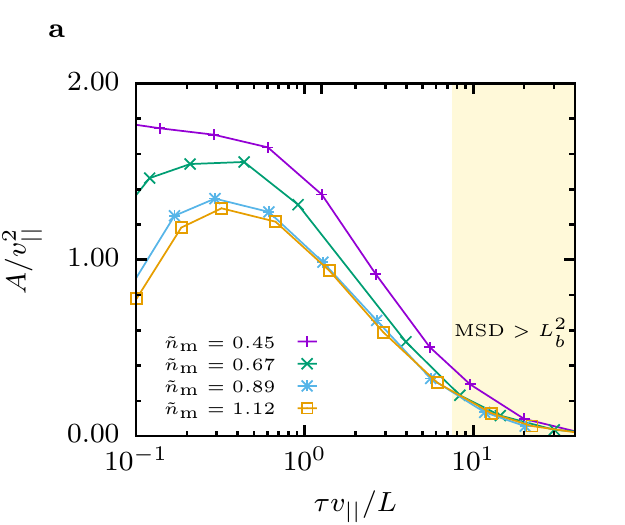}
\includegraphics[width=0.39\textwidth]{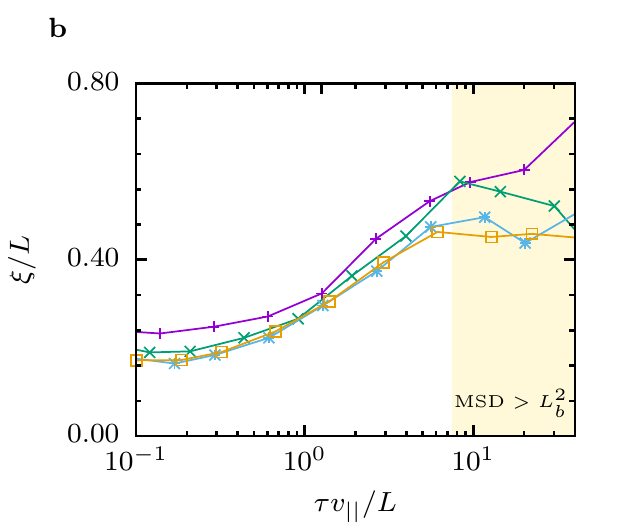}\\
\includegraphics[width=0.39\textwidth]{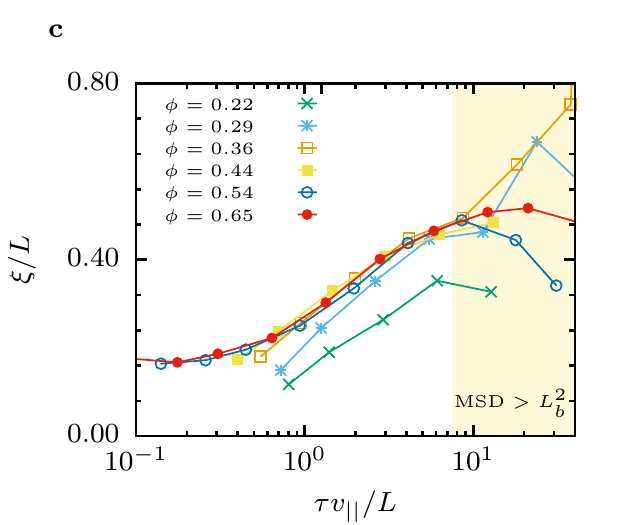}
\caption{{\bf Amplitude and decay length of the velocity correlation
  function as a function the lag time.\\\hspace{\columnwidth}}  
  (\textbf{a},\textbf{b})~Varying the motor density $0.4<\nmt<1.5$, with ($\phi,\ellpt$)=($0.66,3.4$), 
  (\textbf{a})~Amplitude, (\textbf{b}) Velocity decay length. 
  (\textbf{c})~Velocity decay length for different area fractions with $(\ellpt,\nmt)=(3.4,0.89)$.}
\label{fig:dcf_nm}
\end{figure}

Amplitude and velocity correlation length are shown in
\mbox{Fig.~\ref{fig:dcf_nm}\textbf{a} and \textbf{b}}, respectively, for different
activities (changing the number of motors, while keeping all other
parameters constant). Spatial velocity correlations build up at short times and 
decay at long times.  
A maximum amplitude is found at
$\tau v_{\parallel}/L\approx 1/2$, the time at which filaments have moved
about half their length. 
Interestingly, the time-dependent velocity correlation lengths $\xi/L$ for different 
activities superimpose when plotted as a function of $\tau v_{\parallel}/L$. 
The same behaviour of the
velocity correlation length is also found when the area fraction of
filaments is changed as shown in Fig.~\ref{fig:dcf_nm}\textbf{c},
except for very small area fractions ($\phi < 0.3$) where the filaments are still in
the isotropic state (see the snapshots in Fig.~6~\textbf{a},\textbf{b} in
the main text). 

\begin{figure}[h]
\includegraphics[width=0.39\textwidth]{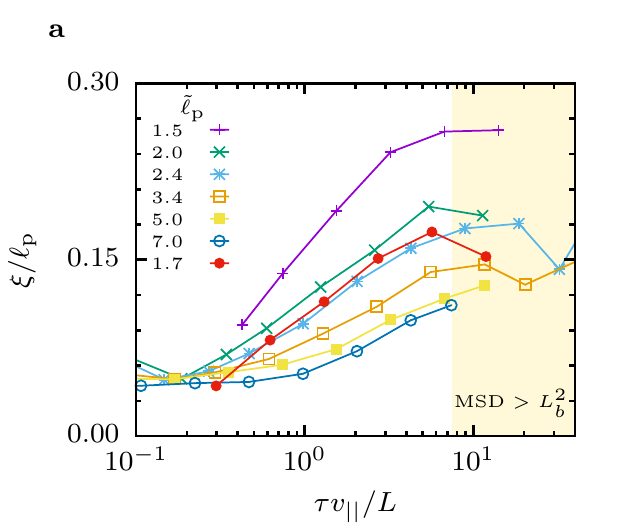}
\includegraphics[width=0.39\textwidth]{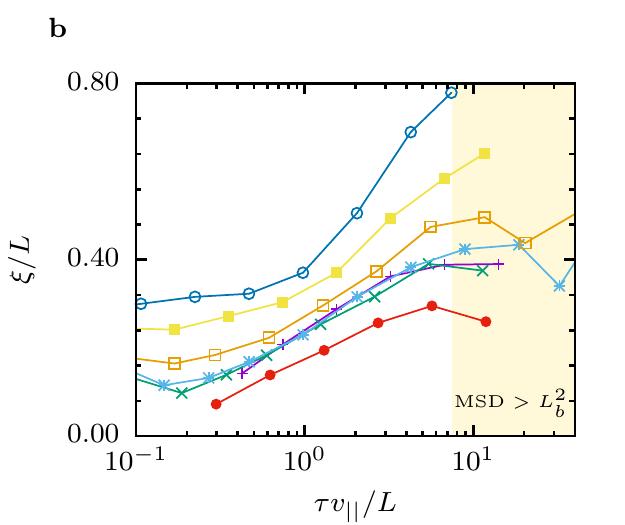}
\caption{{\bf Dependence of the velocity correlation length on the persistence length.\\\hspace{\columnwidth}} ({\textbf a}) Normalisation of $\xi$ by the persistence length $\ellp$, ({\textbf b}) Normalisation of $\xi$ by the filament length $L$. Here all simulations are for $L/\sigma=20$, except for the filled red symbols which are for $L/\sigma=$40.}
\label{fig:dcf_lp}
\end{figure}

However, a different behaviour is found when the filament persistence length is varied,
see Fig.~\ref{fig:dcf_lp}; here the
velocity correlation length $\xi$ increases with 
increasing filament stiffness and saturates to a constant value of approximately $\xi/\ellp\approx 0.04$ for large persistence lengths, see Fig.~\ref{fig:dcf_lp}{\textbf{a}}. 
However, for small persistence lengths $\ellp \approx L$, where domain sizes are small, 
the velocity correlation length is independent of $\ellp$ and approaches a constant $\xi/L=0.16$ for $\tau v_{\parallel}/L=1/2$, see Fig.~\ref{fig:dcf_lp}{\textbf{b}}.
As for most of the results discussed we have focussed on systems with a fixed aspect ratio $L/\sigma=20$. We included simulation data for longer filaments $L/\sigma=40$ which 
show that that the scaling proposed in Fig.~\ref{fig:dcf_lp} reasonable but might be more complicated as the filament thickness is ignored.

Our system is different from the experiments in
Ref.~\cite{SANCHEZ012} where suspensions of extensile bundles of microtubules and kinesin motors are studied and in which the velocity correlations are measured using displacements of large tracer particles. 
It is also different from the continuum model of Ref.~\cite{THAMPI013} based on the theory of  active nematics. 
Nevertheless, the velocity correlation functions are in all cases found to be exponential 
functions, and in particular the correlation length is rather insensitive to the activity. 
Moreover, for larger filament stiffnesses $\ellp \gg L$ we also find an approximately 
linear increase of the velocity correlation length with increasing bending rigidity 
(in our case $\ellpt$) as in active gel theory \cite{THAMPI013}. For small filament 
stiffness our component-based model predicts universal scaling for $\xi/L$ for various 
persistence lengths. 

\noindent{\bf Spatial orientational correlation function}\\
The sizes of the domains can be estimated through the segment-based spatial 
orientational correlation functions, $\Omega_\mathrm{\parallel}(r)$
and $\Omega_\mathrm{\perp}(r)$, which measure
the average angle between filament segments found at a distance $r$ from a
central segment parallel or perpendicular to the orientation of that
central segment, 
\begin{equation*}
\Omega_\mathrm{\parallel}(r)= \left < \frac{\sum\limits_{s,j}  \delta({\hat{{\bf p}}_q^i \cdot \hat{\bf r}_{qs}^{ij}-1}) \delta({|{\bf r}_{qs}^{ij}|-r}) \arccos (\hat{\bf p}_q^i \cdot \hat{\bf p}_s^j)} {\sum\limits_{s,j} \delta({\hat{{\bf p}}_q^i \cdot \hat{\bf r}_{qs}^{ij}-1}) \delta({|{\bf r}_{qs}^{ij}|-r})}\right >_{q,i,t}
\end{equation*}
and
\begin{equation*}
\Omega_\mathrm{\perp}(r)=\left < \frac{\sum\limits_{s,j}  \delta({\hat{{\bf p}}_q^i \cdot \hat{\bf r}_{qs}^{ij}}) \delta({|{\bf r}_{qs}^{ij}|-r}) \arccos (\hat{\bf p}_q^i \cdot \hat{\bf p}_s^j)} {\sum\limits_{s,j} \delta({\hat{{\bf p}}_q^i \cdot \hat{\bf r}_{qs}^{ij}}) \delta({|{\bf r}_{qs}^{ij}|-r})}\right >_{q,i,t} \, .
\end{equation*}
Here, $\sum_{s,j}$ is a short notation for the double sum $\sum_s
\sum_j$; the delta functions select the distances and
orientations. The unit orientation vector of a segment is $\hat{\bf
  p}_q^i=({\bf r}_q^{i+1}-{\bf r}_q^i)/|{\bf r}_q^{i+1}-{\bf r}_q^i|$,
the separation between two segments $i,j$ on filaments $q,s$ is
$r=|{\bf r}_{qs}^{ij}|=|({\bf r}_q^i+{\bf r}_q^{i+1})-({\bf
  r}_s^j+{\bf r}_s^{j+1})|/2$ and $\hat{\bf r}_{qs}^{ij}={\bf
  r}_{qs}^{ij}/|{\bf r}_{qs}^{ij}|$ is the unit separation vector.
 
\begin{figure}[H]
\begin{center}
\includegraphics[width=0.45\textwidth]{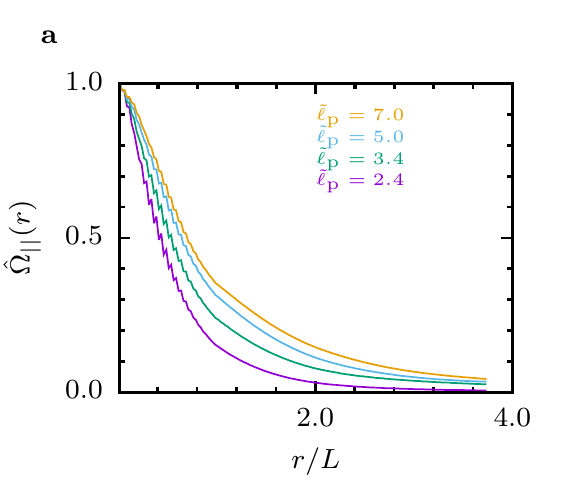}
\includegraphics[width=0.45\textwidth]{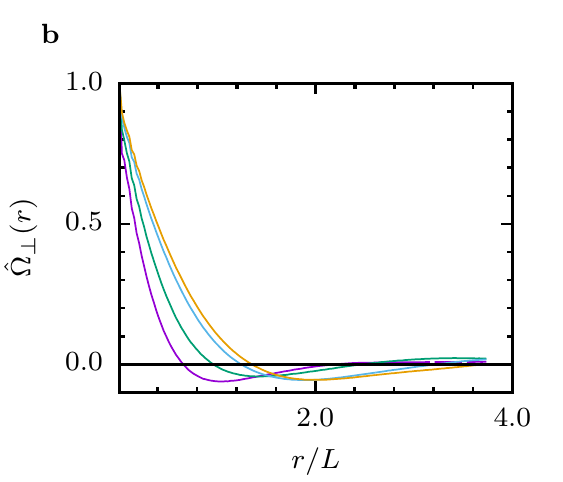}
\caption{{\bf Segment orientational correlation functions.\\\hspace{\columnwidth}} $\textbf{a}$) Longitudinal $\widehat{\Omega}_\mathrm{\parallel}(r)$,  and ($\textbf{b}$) transversal  $\widehat{\Omega}_\mathrm{\perp}(r)$ for $(\phi, \nmt) = (0.66,0.89)$.}
\label{fig:omega}
\end{center}
\end{figure}

The functions $\Omega_\mathrm{\parallel}(r)$ and
$\Omega_\mathrm{\perp}(r)$ are zero for small distances and saturate
to $\pi/2$ at large distances, where parallel and
antiparallel orientations are equally probable. Therefore, the displaced and
normalised functions $\widehat{\Omega}(r)=1-2\Omega(r)/\pi$ decay rapidly as shown in Fig.~\ref{fig:omega}. 

\onecolumngrid

This decay allows the
estimation of  the domain sizes parallel and perpendicular to the filament orientation.
The length $l_\parallel$ and the width $l_\perp$ are defined as twice
the peak width at half maximum. The examples of $\widehat{\Omega}(r)$ displayed in
Fig.~\ref{fig:omega} show that the decay length becomes smaller when
the persistence length decreases, i.e., domains become smaller. The
transversal orientational correlation function shows oscillations, indicating the
abrupt change of the orientation when the interface
between domains is crossed.


\bibliographystyle{apsrev4-1}
\bibliography{filmot}